\documentclass{aa}
\pdfoutput=1
\usepackage{amssymb}
\usepackage{natbib}
\usepackage{mathtools}
\usepackage[version=3]{mhchem}
\usepackage{siunitx}
\usepackage{multirow}
\usepackage{lipsum}
\usepackage{blindtext}
\usepackage{array}
\usepackage{subcaption}
\usepackage{fixltx2e}
\usepackage{tablefootnote}
\usepackage{epstopdf}
\newcolumntype{H}{>{\setbox0=\hbox\bgroup}c<{\egroup}@{}}

\begin{document}
\title{Mapping CS in starburst galaxies:\\ Disentangling and characterising dense gas}

\author{G. Kelly\inst{\ref{inst1}}\and S. Viti\inst{\ref{inst1}}\and E. Bayet\inst{\ref{inst2}}\and R. Aladro\inst{\ref{inst3}}\and J. Yates\inst{\ref{inst1}}}

\institute{Department of Physics and Astronomy, University College London, Gower Street, London, WC1E 6BT, U.K. email{: g.kelly@ucl.ac.uk}\label{inst1} \and
Sub-department of Astrophysics, University of Oxford, U.K.\label{inst2} \and
European Southern Observatory, Santiago, Chile\label{inst3}}



\abstract{}{We observe the dense gas tracer CS in two nearby starburst galaxies to determine how the conditions of the dense gas varies across the circumnuclear regions in starburst galaxies.}{Using the IRAM-30m telescope, we mapped the distribution of the CS$(2-1)$ and CS$(3-2)$ lines in the circumnuclear regions of the nearby starburst galaxies NGC\,3079 and NGC\,6946. We also detected formaldehyde (H$_{2}$CO) and methanol (CH$_{3}$OH) in both galaxies. We marginally detect the isotopologue C$^{34}$S.}  {We calculate column densities under LTE conditions for CS and CH$_{3}$OH. Using the detections accumulated here to guide our inputs, we link a time and depth dependent chemical model with a molecular line radiative transfer model; we reproduce the observations, showing how conditions where CS is present are likely to vary away from the galactic centres.}{{Using the rotational diagram method for CH$_{3}$OH, we obtain a lower limit temperature of 14 K. In addition to this, by comparing the chemical and radiative transfer models to observations, we determine the properties of the dense gas as traced by CS (and CH$_{3}$OH). We also estimate the quantity of the dense gas.} We find that, {provided there are between \num{e5} and \num{e6} dense cores in our beam}, for both target galaxies, emission of CS from warm (T = 100 - 400 K), dense (n(H$_{2}$) = 10$^{5-6}$ cm$^{-3}$) cores, possibly with a high cosmic ray ionisation rate ($\zeta$ = 100 $\zeta_{0}$) best describes conditions for our central pointing. In NGC 6946, conditions are generally cooler and/or less dense further from the centre, whereas in NGC 3079, conditions are more uniform. The inclusion of shocks allows for more efficient CS formation, which means that gas that is less dense by an order of magnitude is required to replicate observations in some cases.} 

\maketitle
\section{Introduction}
The formation of stars, in particular the most massive ones, is one of the most important processes regulating the evolution
of galaxies;
to trace this process, it is crucial to identify the chemical nature of the gas involved
and to determine the physical conditions that are more prone to lead to star formation.
Massive stars usually form in large
concentrations of  dense, warm gas. These conditions are able to survive disruptive winds and radiation
from nearby newly formed stars  \citep{2007ARA&A..45..481Z}. In particular, starburst galaxies are powered by exceptionally high
rates of massive star formation, possibly triggered by mergers between
galaxies rich in interstellar matter. Such events must be relatively
short-lived, persisting only until the interstellar gas reservoir is
significantly depleted.

Subject to a proper interpretation, observations of molecules in starbursts can be used for many purposes: tracing the leftover reservoir of matter from the star formation process; tracing the star formation process
itself; and determining the galaxy energetics through the influence of young stars on the surrounding environments \citep{2014ApJ...780L..13K, 2014ApJ...796L..15S}. 
More specifically, the detection of molecular star-forming gas is one of the most direct ways to measure the star formation rate and activity in a galaxy, allowing us to significantly improve our understanding of galaxy formation and evolution. To date, the wealth of molecular data for at least the nearest galaxies,
shows a
chemical
diversity and complexity that cannot be explained by a
one-component, static model, and indicates how relative abundances between
molecules may be able to provide insights into the physical distribution
of the molecular gas and the energetics of these galaxies.

While some observed molecular transitions may arise from a UV-dominated gas, many
can only be explained by the presence of large reservoirs of dense, warm gas (possibly pre-processed on dust grains) that are a part of the star formation process \citep{2004ApJ...606..271G, 1998ApJ...502..296O}. 
In dense regions (n(H$_{2}$ \textgreater \num{e4} cm$^{-3}$), the
galactic interstellar radiation field is unimportant for the chemistry
since it occurs in conditions of high visual extinction.
Molecules that are particularly enhanced during the star formation process (including the warm up phase that follows the birth of protostars)
are therefore good tracers of the
formation of massive stars in galaxies. 

From a chemical point of view,
sulphur-bearing species have been extensively shown to be particularly enhanced during massive star formation in our own Galaxy, as well as in nearby galaxies \citep{1989A&A...226L...5M, 2006ApJS..164..450M}. 
Among sulfur-bearing species, carbon monosulphide (CS) appears as one of the best tracers of very dense gas \citep{1996A&AS..115...81B}.
Observationally, CS has been identified in several studies of nearby galaxies \citep{1989A&A...226L...5M, 2006ApJS..164..450M, 2008ApJ...685L..35B, 2011A&A...525A..89A}. Following theoretical studies \citep{2008ApJ...676..978B, 2011ApJ...728..114B}, multi-line observations of the 
CS molecule in nearby (D $<$ 10 Mpc) extragalactic environments were carried out \citep{2009ApJ...707..126B}.  These studies show that despite the lack of spatial resolution it is possible to point
to a reservoir of gas of relatively high density, $\sim$ 10$^5$ cm$^{-3}$, traced by the $low$ $J$ (up to 4) transitions of CS 
and an even higher ($>$ 10$^6$ cm$^{-3}$) density component traced by the $high$ $J$ (up to 7)
transitions. So far surveys have focused on CS detections in the centre of galaxies. To gain insight into the distribution of this gas in the environment surrounding the nucleus, galaxy mapping in CS was required.

\begin{figure}[htbp]
\begin{center}

\includegraphics[width=0.92\linewidth, angle=0]{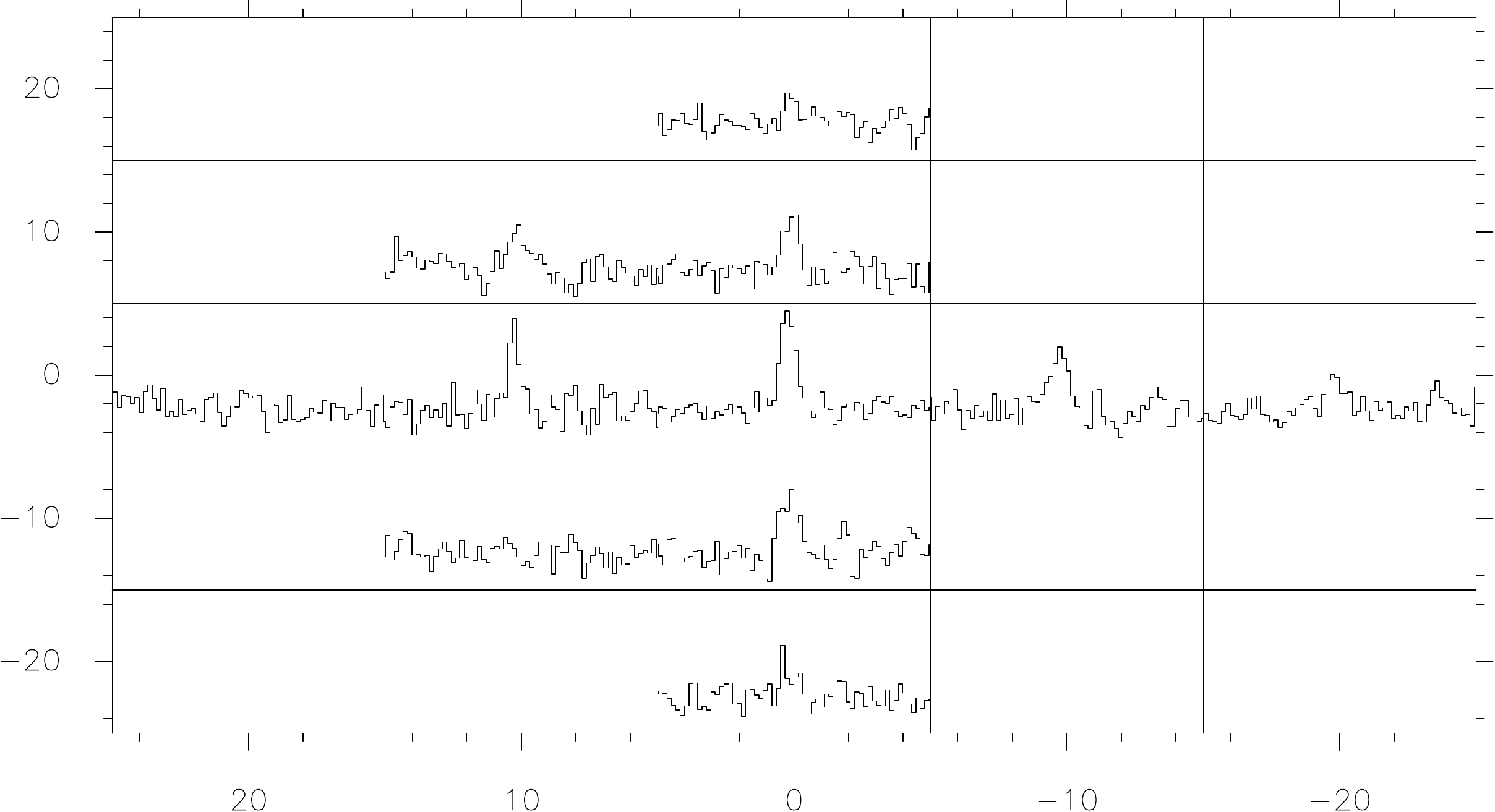}
\includegraphics[width=0.5\linewidth, angle=270]{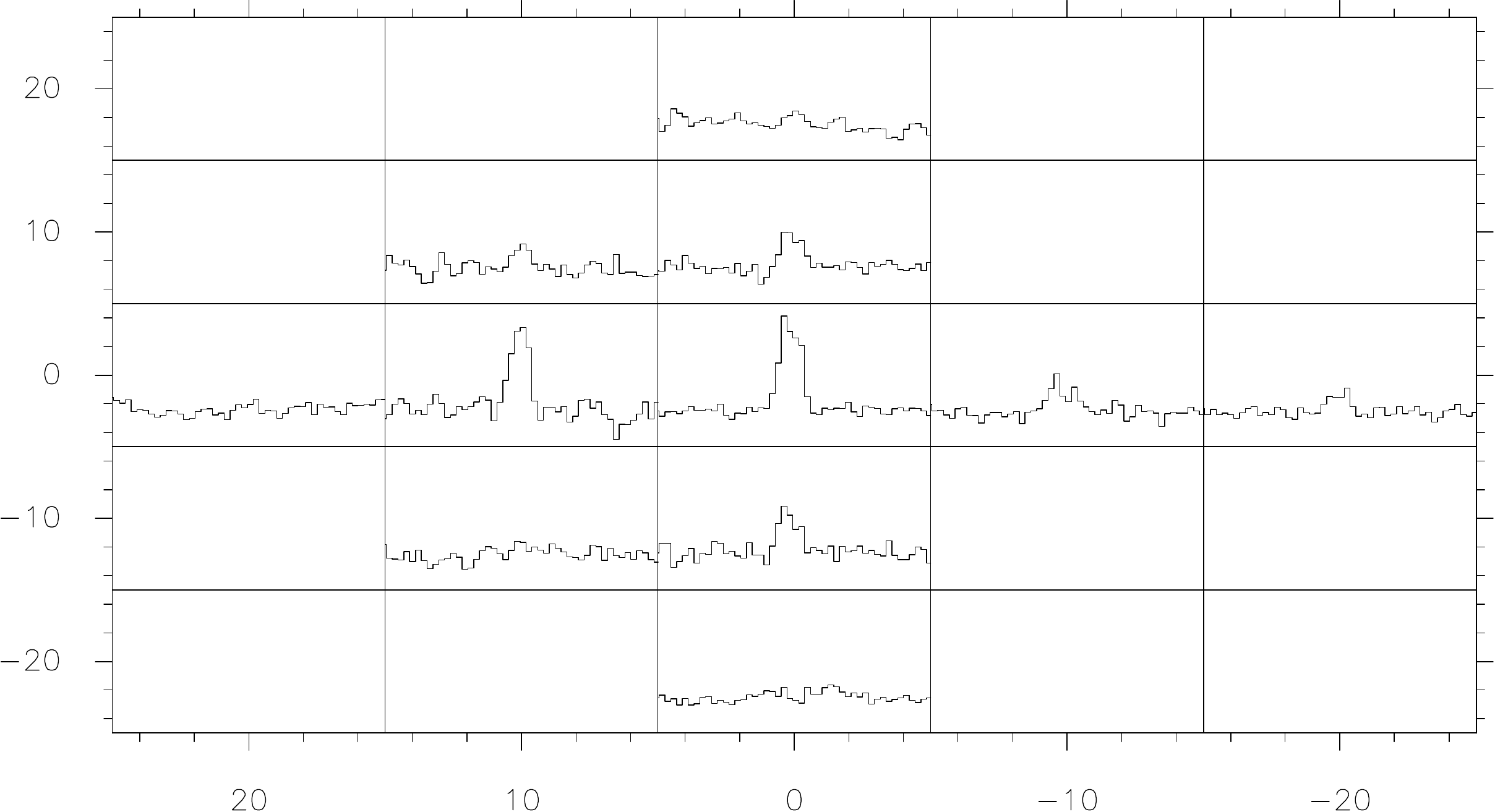}
\includegraphics[width=0.15\linewidth, angle=270]{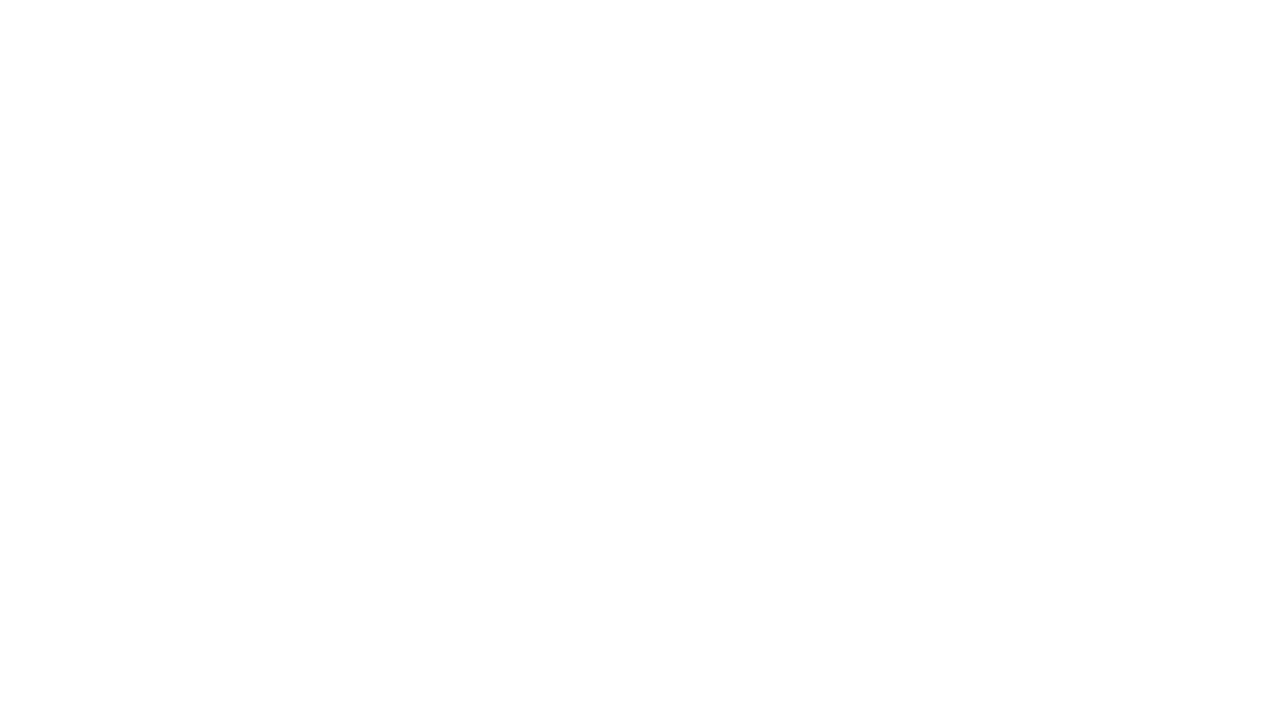}
\includegraphics[width=0.5\linewidth, angle=270]{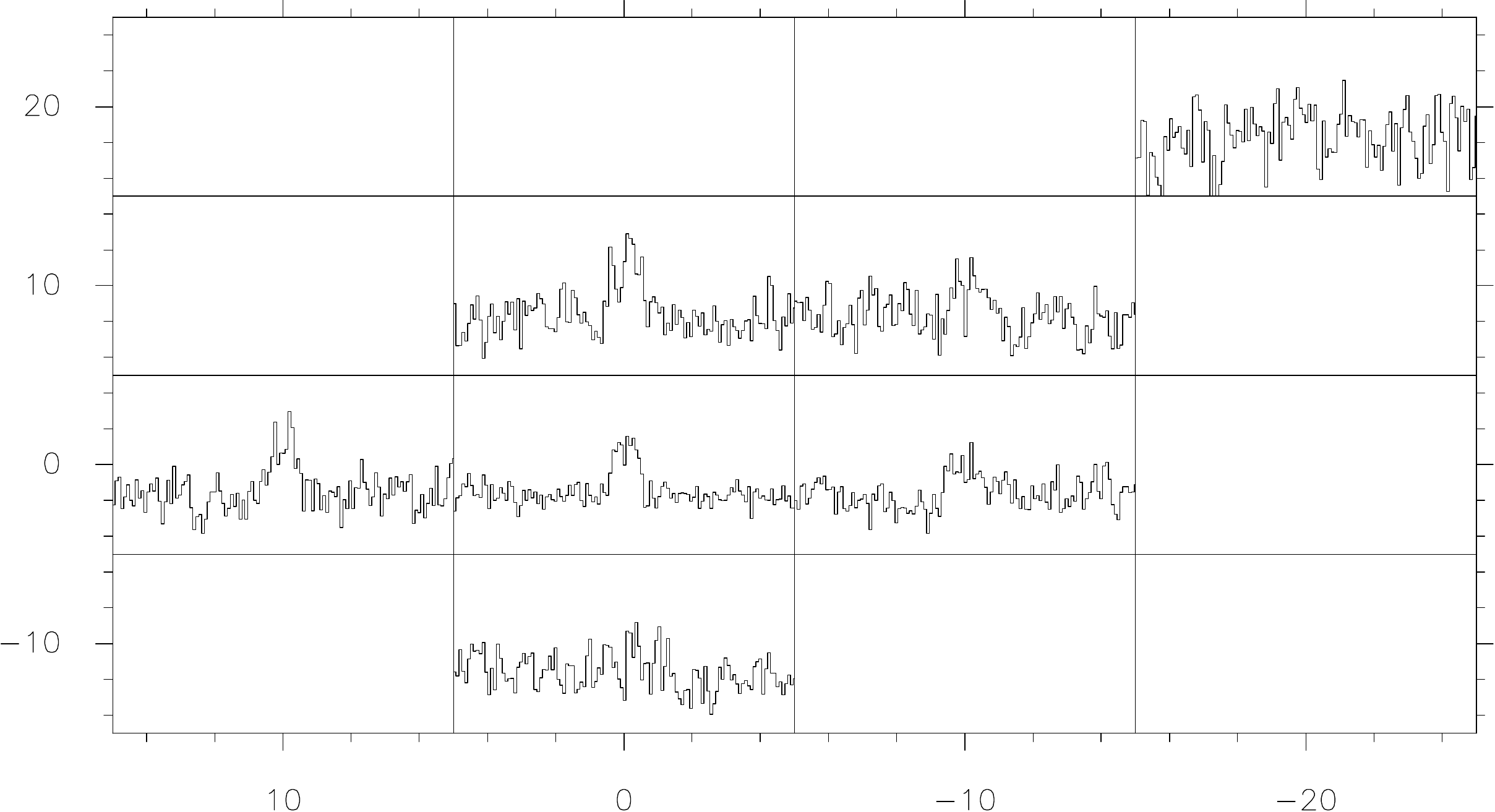}
\includegraphics[width=0.5\linewidth, angle=270]{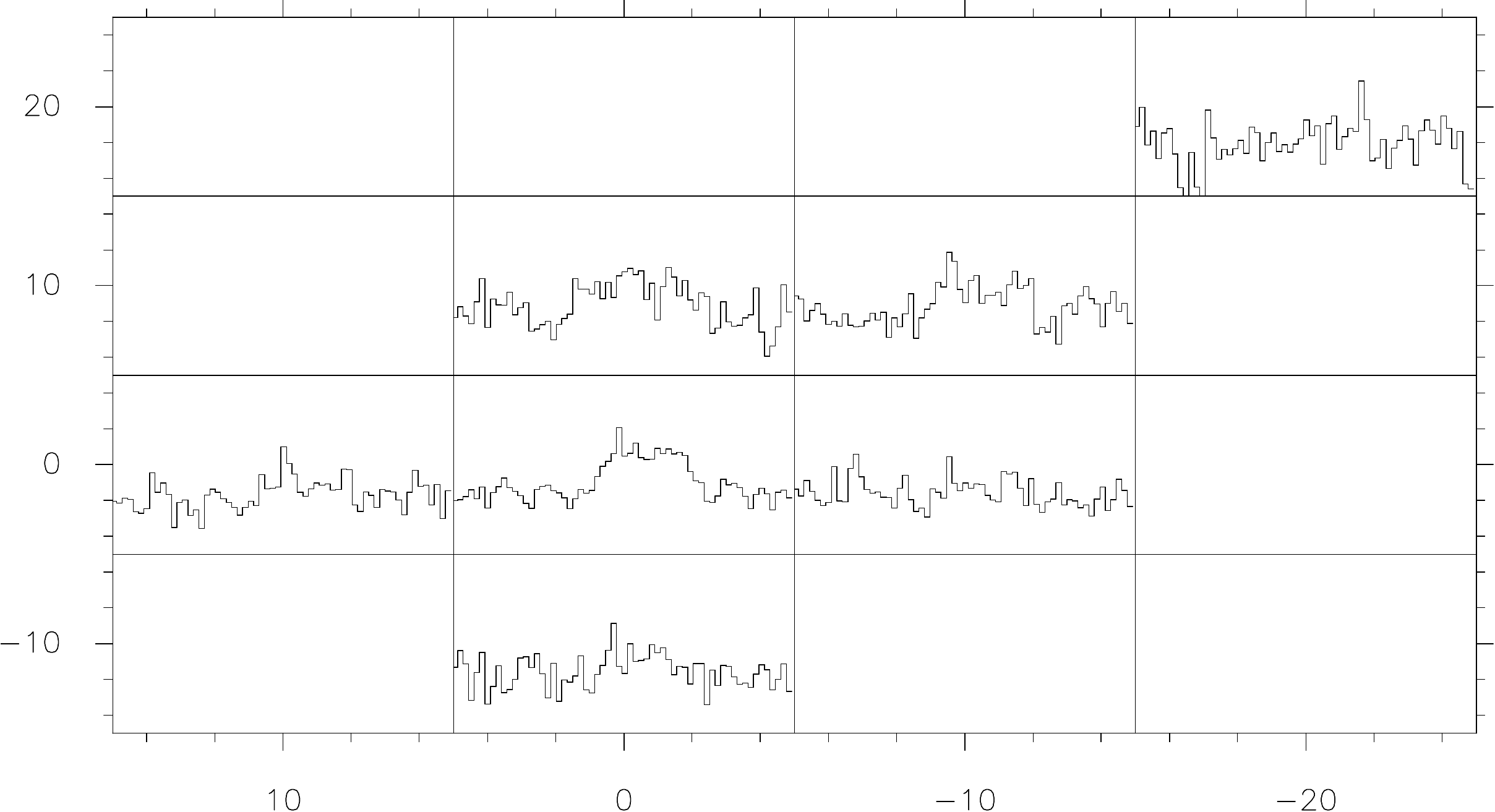}

\caption{Maps of NGC 6946 in CS (J = 2 - 1) and CS (J = 3 - 2) [top two] and the same for NGC 3079 [bottom two]. Units are arc seconds. $\Delta$v = 8 km s$^{-1}$}
\label{fig:maps}
\end{center}
\end{figure}

\section{Target galaxies}

In this work, the spatial distribution of CS in two nearby galaxies, NGC 6946 and NGC 3079, is studied. We detect CS $J$ = $(2-1)$ and $J$ = $(3-2)$ lines in the central region and various offset positions in each galaxy. NGC 3079 is a nearby Seyfert 2 galaxy with a high inclination angle of 84 degrees and a nuclear bubble of starburst activity \citep{1991ApJ...371..111I,1995ApJ...446..602B}. NGC 6946 is a nearby late type spiral galaxy with numerous registered supernovae, also with nuclear starburst activity \citep{1996ApJ...467..227E}. We observe CS lines in the central point of each galaxy as well as at offset points staggered at increments of 10'' (Figure ~\ref{fig:maps}). At a distance of $\approx$19 Mpc and $\approx$5.5 Mpc , we observe 92 pc arcsec$^{-1}$ and 27 pc arcsec$^{-1}$ in NGC 3079 and NGC 6946, respectively. As a consequence, for NGC 3079 we map out to a maximum of 2.8 kpc from our central pointing, while in NGC 6946 we map a smaller region, 810 pc. Further information on both galaxies can be found in Table ~\ref{tab:galaxies}.

\citet{1994ApJ...433...48V} discovered the line emitting superbubble in NGC 3079 extending 13'' to the east of the nucleus. Within this bubble, they find extremely violent gas motions ranging over 2000 km s$^{-1}$, which is also seen to the opposite, western side of the nucleus. In addition to this, unusual excitation shows that shocks may be important and could be a significant contributor to line emission across the bubble. From CO observations, \citet{2002ApJ...573..105K} show that the central 2 kpc region of the galaxy contains a smoothly distributed $\approx$ \num{2e9} M$_{\odot}$ of molecular gas. Also using CO observations, \citet{2011AAS...21724614L} observe that NGC 6946 is also highly rich in molecular gas, both in the central region we observe and out into the details of the spiral arms. Starburst activity in the nucleus of NGC 6946 is more moderate than NGC 3079 \citep{1983ApJ...268L..79T}. NGC 6946 has a small 15'' bulge, within a larger 63'' bar around it. Dense gas has been studied in the central region of this galaxy using tracers HCN \citep{2007A&A...462L..27S} and HCO$^{+}$ \citep{2008ApJ...673..183L}. \citet{2008ApJ...673..183L} study the central kpc and find that the density of the gas is roughly constant across the area studied and find that a plausible distribution is of dense clumps bound by self-gravity, interspersed throughout the lower density molecular gas. However, they also state that it could be possible that the dense gas is homogeneously distributed about the galactic centre. {\citet{2007A&A...462L..27S} observed in great detail the central 50 pc of the galaxy and find that the HCN$(1-0)$ intensity distribution in the very centre of the galaxy is centred around two peaks. They find that these peaks correspond quite well to the molecular gas as traced by CO$(2-1)$. They state that one possible interpretation is that these gas peaks are the result of two spiral arms connecting in the central 100 pc.} Although there have been detections of CS in the centre of many nearby galaxies (e.g. \citealt{2014ApJ...784L..31Z}), a mapping and full analysis of its distribution in those galaxies has not yet been undertaken. The location and the characteristics of this gas away from galactic centres is crucial to our understanding of the processes and conditions that a necessary for massive star formation to occur.

\begin{table*}[hbtp]
\begin{center}
\caption{Galaxy properties}
\label{tab:galaxies}
\begin{tabular}{c c c c c c}

Galaxy & RA (J2000) & Dec (J2000) & Velocity (km s$^{-1}$) & Classification & SFR (M$_{\odot}$ yr$^{-1}$)\\
\hline
NGC 6946 & 20:34:51.88 & 60:09:14.9 & 48$^{a}$ & Scd & 3.2$^{b}$  \\
NGC 3079 & 10:01:57.79 & 55.40:47.0 & 1124$^{a}$ & SB(s)c  & 6$^{c}$\\

\end{tabular}
\caption*{\tiny NOTES. a) \citet{2007ApJ...655..790C}, b) \citet{2008AJ....136.2782L}, c) \citet{2014MNRAS.437.1698M}}
\end{center}
\end{table*}

\section{Observations}
The observations were carried out in August 2010 using the IRAM-30m telescope to include the CS$(2-1)$ and CS$(3-2)$ lines ($\nu$ = 97.980 GHz and 146.969 GHz, respectively). In order to capture other species (methanol, formaldehyde) and isotopologues of CS, each observation was not centred upon the CS line. Instead a line survey from 95.5 GHz to 99 GHz and from 144 GHz to 147.5 GHz was performed in order to detect as many chemical tracers as possible of the dense, star-forming gas. At the observed frequencies of CS$(2-1)$ and CS$(3-2)$, IRAM 30m has a HPBW of $\approx$ 25$''$ and 17$''$ with main beam efficiencies, $B_{\rm eff}$, of 0.81 and 0.74, respectively. {These beam sizes correspond to between approximately 450 pc and 650 pc (NGC 6946), and 1.5 kpc and 2.3 kpc (NGC 3079).}. During observing, approximately every 2 hours, the pointing, focus and calibration were completed on planets and evolved stars. {The pointing error was estimated to be $\le$ 3$''$.} The observations were carried out under varying weather conditions. $T_{\rm sys}$ was between 100 K and 300 K with medium-good weather conditions ($\tau_{225}$ \textless 0.15). The observations were done by wobbling the secondary mirror with a beam throw of 180$''$ in azimuth.  For the observed spectra, the antenna temperature (T$_{a}$*) was converted to main beam temperature ($T_{\rm mb}$) using $T_{\rm mb}$ = ($F_{\rm eff}/B_{\rm eff}$)$T_{\rm a}$*, where $F_{\rm eff}$ is the forward efficiency of the telescope (values ranging from 0.93 to 0.95). The channel width spacing was 4 km s$^{-1}$ and has been smoothed down to a 8 km s$^{-1}$. The EMIR E0 and E1 receivers were used in configurations to optimise observing time. We combined this with the Wilma backend. The data reduction was carried out using the CLASS program, part of GILDAS software\footnote{http://www.iram.fr/IRAMFR/GILDAS}.

\begin{figure*}[htbp]
\begin{center}
\includegraphics[width=0.4\linewidth,angle=270]{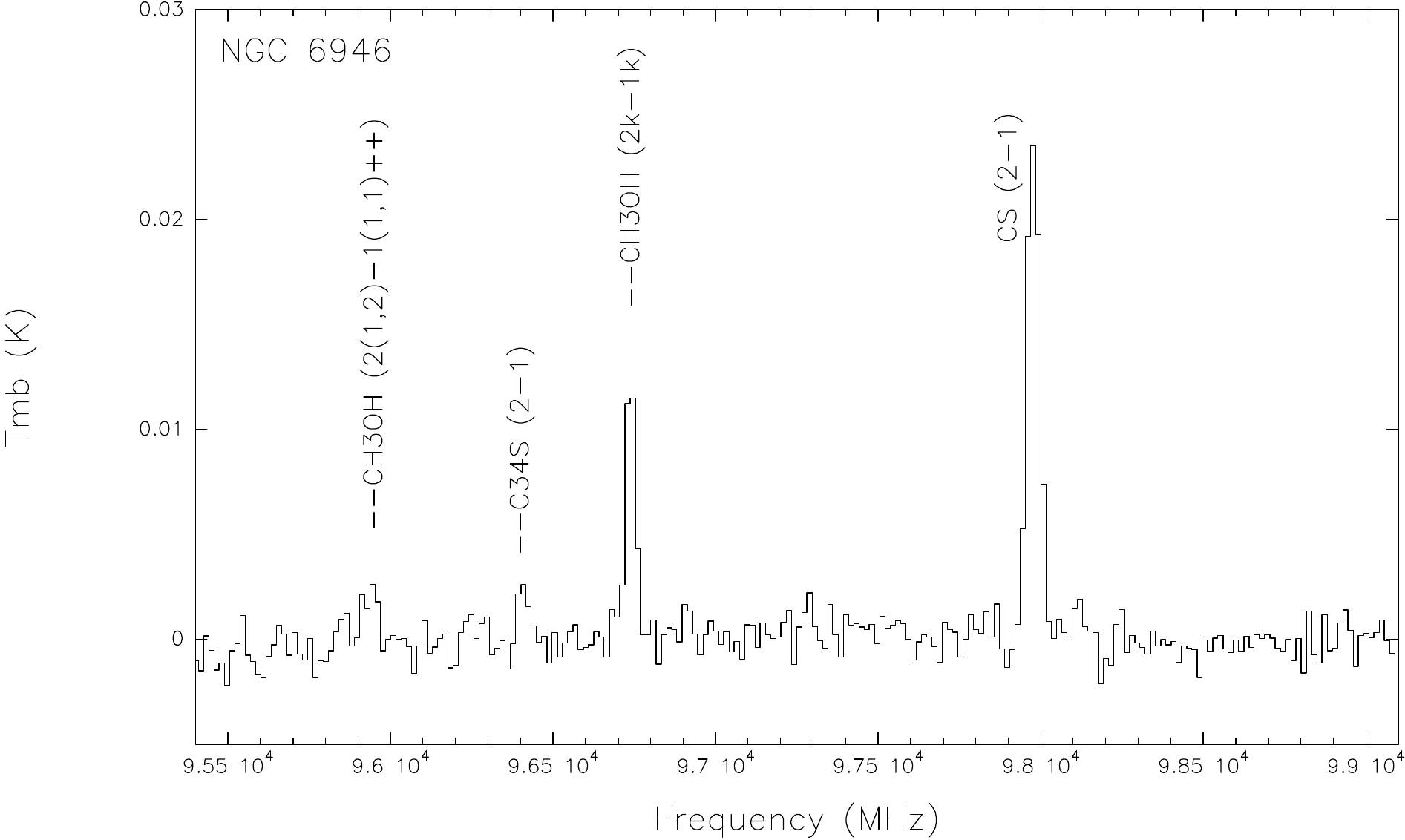}
\includegraphics[width=0.4\linewidth,angle=270]{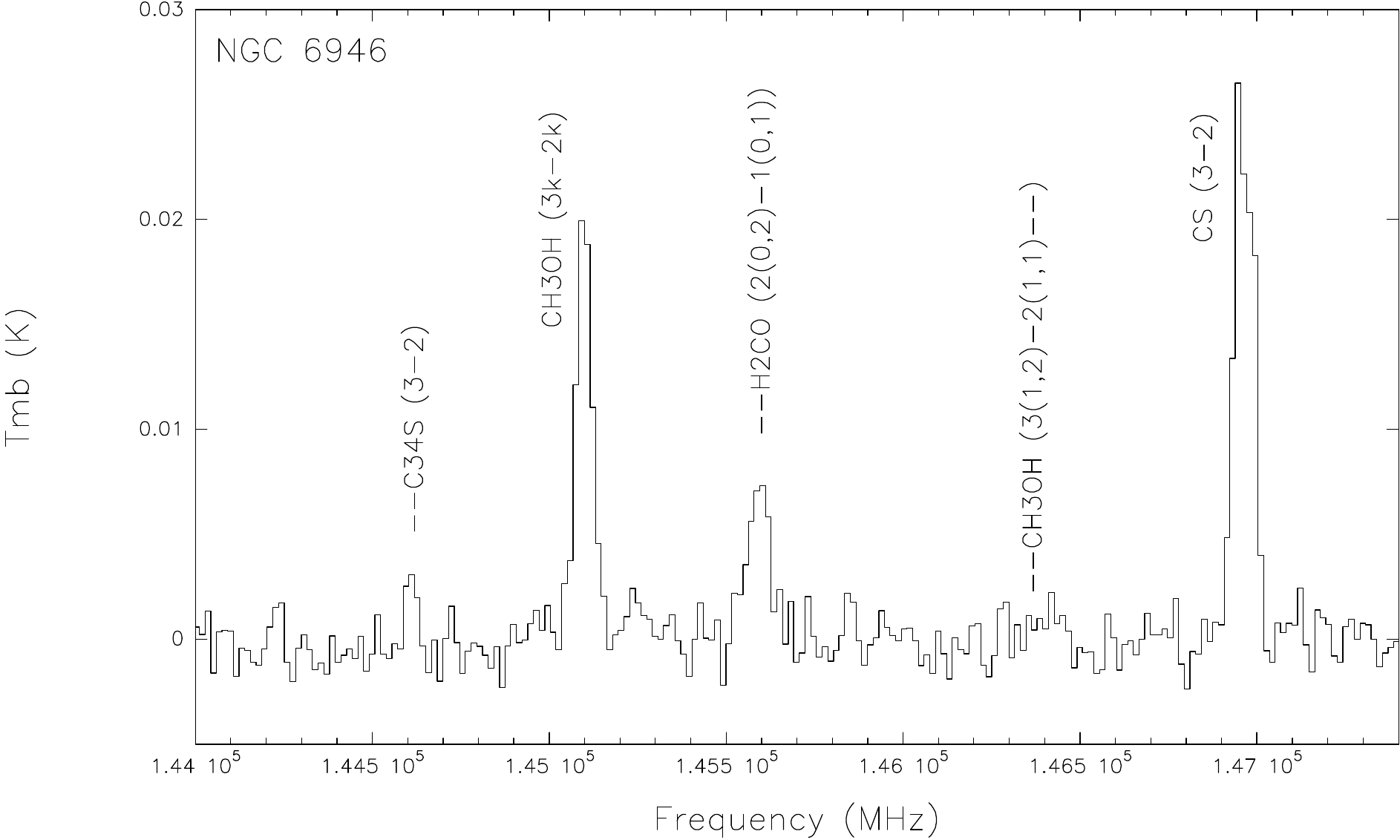}

\caption{Potential {detections} in centre of NGC 6946}
\label{fig:detections}
\end{center}
\end{figure*}

\begin{figure*}[htbp]
\begin{center}
\includegraphics[width=0.4\linewidth,angle=270]{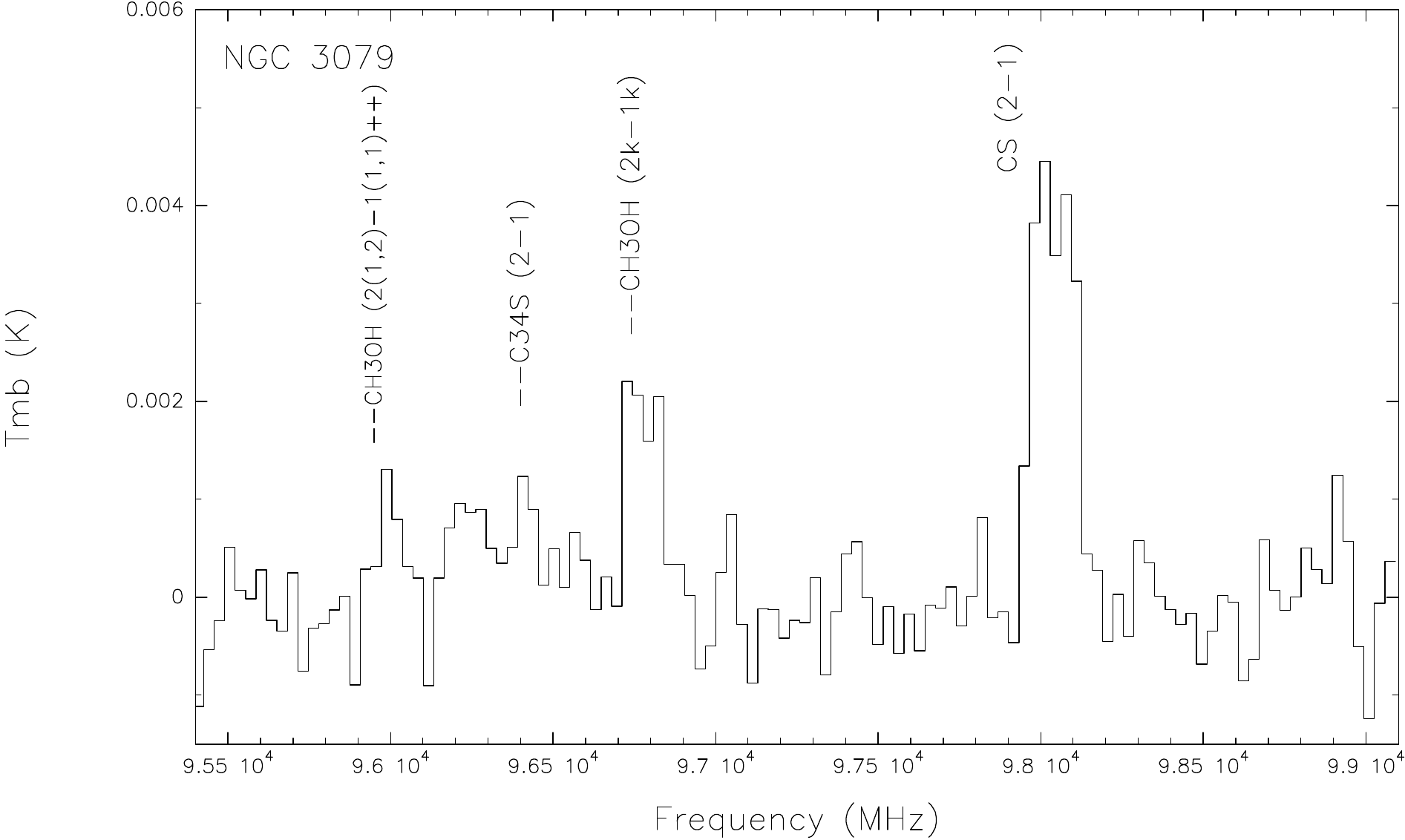}
\includegraphics[width=0.4\linewidth,angle=270]{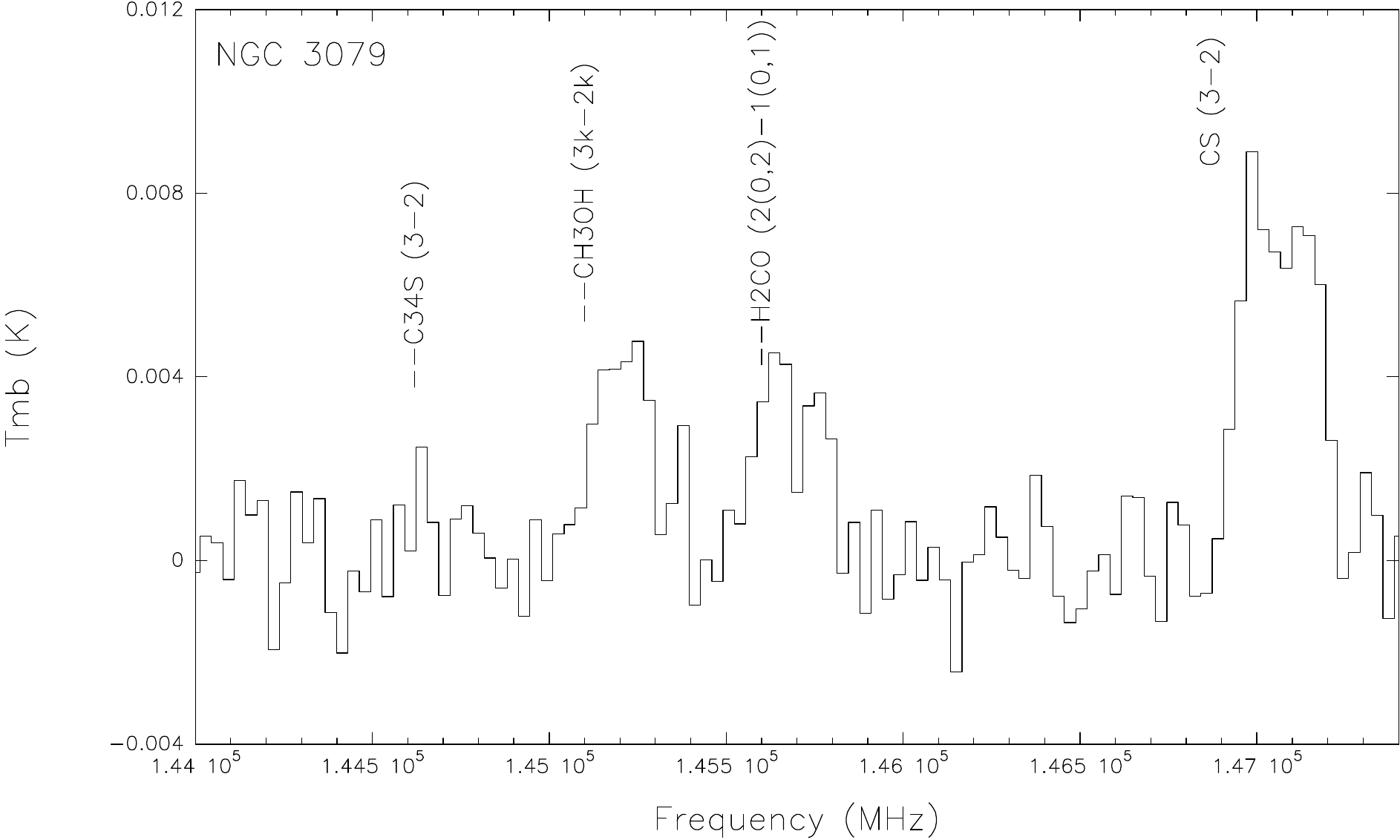}

\caption{Potential detections in centre of NGC3079}
\label{fig:detections}
\end{center}
\end{figure*}

\begin{table*}[htbp]
\scriptsize
\begin{center}
\caption{NGC 6946 observational parameters and Gaussian fits. For a two component fit, NGC 6946-1 corresponds to the component at lower velocity and NGC 6946-2 the higher velocity component.}
\label{tab:6946obs}
\begin{tabular}{ c r r H H r r H H r r r r r}

Source & Line & $\nu$ & RA (J2000) & Dec (J2000) & Offset & Offset  & T$_{sys}$ & Beam & $\int$(T$_{mb}$d$\nu$) & FWHM & T$_{peak}$ & rms & v$_{peak}$ \\    
 & & (GHz) & (h:m:s) & ( $^{\circ}$ : ' : " )  & RA (") & Dec (") & (K) & Size (") & (K km s$^{-1}$) & (km s$^{-1}$) & (mK) & (mK) & (km s$^{-1}$) \\
\hline\hline
NGC 6946 & CS$(2-1)$ & 97.980 & 20:34:51.88 & 60:09:14.9 & 0 & 0 & 133 & 25.1 & 3.9 $\pm$ 0.1 &140.1 $\pm$ 4.2 & 25.9 & 1.6 & 60.4 $\pm$ 1.7\\
 & &  &  &  &  0 & -10 & 132 & " & 2.2 $\pm$ 0.1 &136.6 $\pm$ 8.6 & 15.0 & 2.1 & 69.0 $\pm$ 4.1 \\     
 & &  &  &  &  0 & +10 & 131 & " & 1.9 $\pm$ 0.1 &146.8 $\pm$ 10.9 & 12.2 & 2.1 & 56.1 $\pm$ 4.7 \\ 
 & &  &  &  &  +10 & 0 & 140 & " & 2.9 $\pm$ 0.1 &139.2 $\pm$ 7.3 & 19.8 & 2.5 & 47.0 $\pm$ 3.6 \\ 
 & &  &  &  &  -10 & 0 & 127 & " & 2.1 $\pm$ 0.1 &133.8 $\pm$ 7.0 & 14.8 & 2.0 & 70.4 $\pm$ 3.5 \\ 
 & &  &  &  &  -20 & 0 & 113 & " & 1.6 $\pm$ 0.1 &157.6 $\pm$ 13.7 & 9.3 & 1.7& 70.3 $\pm$ 6.4  \\  
 & &  &  &  &  +20 & 0 & 104 & " & 1.0 $\pm$ 0.1 &150.1 $\pm$ 18.1 & 6.3 & 2.2 &  33.9 $\pm$ 10.3 \\   
 & &  &  &  &  +10 & +10 & 130 & " & 1.4 $\pm$ 0.1 &140.0 $\pm$ 15.7 & 9.1 & 2.3 & 41.6 $\pm$ 6.2 \\
 & &  &  &  &  +10 & -10 & 125 & " & - & - & - & - &  - \\
 & &  &  &  &  0 & -20 & 117 & " & 1.1 $\pm$ 0.1 &123.5 $\pm$ 15.3 & 8.6 & 1.9 & 71.8 $\pm$ 6.6  \\
 & &  &  &  &  0 & +20 & 114 & " & 0.9 $\pm$ 0.1 &126.0 $\pm$ 22.4 & 6.4 & 2.0 & 35.0 $\pm$ 10.1  \\
 & CS$(3-2)$ & 146.969 & 20:34:51.88 & 60:09:14.9 & 0 & 0 & 189 & 16.7 & 3.8 $\pm$ 0.1 &135.2 $\pm$ 3.7 & 26.3 & 2.1 & 63.3 $\pm$ 1.8 \\ 
 & &  &  &  &  0 & -10 & 271 & " & 1.7 $\pm$ 0.2 &124.2 $\pm$ 16.5 & 12.7 & 3.3 & 74.0 $\pm$ 6.7 \\ 
  & &  &  &  &  0 & +10 & 209 & " & 1.4 $\pm$ 0.2 & 128.5 $\pm$ 15.8 & 10.3 & 3.9 & 54.1 $\pm$ 7.7 \\ 
 & &  &  &  &  +10 & 0 & 230 & " & 3.1 $\pm$ 0.2 &114.6 $\pm$ 8.8 & 25.5 & 4.9 & 48.1 $\pm$ 4.1 \\
 & &  &  &  &  -10 & 0 & 202 & " & 1.2 $\pm$ 0.2 & 160.8 $\pm$ 7.2 & 7.3 & 3.4 & 79.7 $\pm$ 11.4 \\ 
  & &  &  &  &  -20 & 0 & 156 & " & 0.7 $\pm$ 0.1 & 124.3 $\pm$ 20.1 & 5.3 & 2.8 & 38.4 $\pm$ 11.3 \\    
 & &  &  &  &  +20 & 0 & 134 & " &  - & - & - & - & - \\  
 & &  &  &  &  +10 & +10 & 205 & " & 0.7  $\pm$ 0.1 & 98.1 $\pm$ 19.8 & 6.9 & 3.5 & 35.4 $\pm$ 11.0\\  
 & &  &  &  &  +10 & -10 & 196 & " & - & - & - & - & -\\  
 & &  &  &  &  0 & -20 & 166 & " & - & - & - & - & -\\   
 & &  &  &  &  0 & +20 & 161 & " & 0.4 $\pm$ 0.1 & 93.6 $\pm$ 29.8 & 3.9 & 3.0 & 32.9 $\pm$ 15.1 \\

 & C$^{34}$S$(2-1)$ & 96.413 &  20:34:51.88 & 60:09:14.9   &  0 & 0 & 133 & 25.1 & 0.4 $\pm$ 0.1 & 122.5 $\pm$ 25.9 & 3.3 & 1.4 & 65.9 $\pm$ 12.7 \\  
 
 & C$^{34}$S$(3-2)$ & 144.617 &  20:34:51.88 & 60:09:14.9   &  0 & 0 & 189 & 16.7 & 0.3 $\pm$ 0.1 & 68.3 $\pm$ 16.2 & 3.6 & 2.1 & 64.7 $\pm$ 9.8 \\  
 
 & H$_{2}$CO(2(0,2)-1(0,1)) & 145.603 &  20:34:51.88 & 60:09:14.9   &  0 & 0 & 189 & 16.7 & 1.3 $\pm$ 0.1 & 161.2 $\pm$ 16.4 & 7.6 & 2.0 & 73.0 $\pm$ 6.2 \\

 & CH$_{3}$OH($2_{k}-1_{k}$) & 96.745 &  20:34:51.88 & 60:09:14.9  &  0 & 0 & 133 & 25.1 & 1.5 $\pm$ 0.1 & 97.8 $\pm$ 6.2 & 14.5 & 1.6 & 71.5 $\pm$ 2.5 \\  

 & CH$_{3}$OH($3_{k}-2_{k}$) & 145.125 &  20:34:51.88 & 60:09:14.9   &  0 & 0 & 189 & 16.7 & 2.4 $\pm$ 0.1 & 106.0 $\pm$ 5.0 & 21.4 & 2.1 & 105.3 $\pm$ 1.8  \\ 
 
 \hline
 NGC 6946-1 & CS$(2-1)$ & 97.980 & 20:34:51.88 & 60:09:14.9 & 0 & 0 & 133 & 25.1 & 1.25 $\pm$ 0.2 & 65.1 $\pm$ 7.2 & 18.0 & 2.1 & 45.1 $\pm$ 3.3 \\
 & &  &  &  &  0 & -10 & 132 & " & 0.6 $\pm$ 0.1 &49.9 $\pm$ 10.0 & 11.3 & 2.1 & -0.8 $\pm$ 5.1 \\     
 & &  &  &  &  0 & +10 & 131 & " &  - & - & - & - & -\\  
 & &  &  &  &  +10 & 0 & 140 & " & 1.23 $\pm$ 0.2 &61.1 $\pm$ 9.6 & 19.1 & 2.5 & 16.5 $\pm$ 5.3 \\ 
 & &  &  &  &  -10 & 0 & 127 & " & 0.68 $\pm$ 0.2 &59.5 $\pm$ 9.4 & 10.7 & 2.0 & 12.8 $\pm$ 5.2 \\ 
 & &  &  &  &  -20 & 0 & 113 & " &  - & - & - & - & -\\  
 & &  &  &  &  +20 & 0 & 104 & " & 0.4 $\pm$ 0.1 & 42.7 $\pm$ 9.5 & 9.7 & 2.2 & -5.7 $\pm$ 5.0 \\   
 & &  &  &  &  +10 & +10 & 130 & " &  - & - & - & - & -\\  
 & &  &  &  &  +10 & -10 & 125 & " &  - & - & - & -& -\\  
 & &  &  &  &  0 & -20 & 117 & " &  - & - & - & - & -\\  
 & &  &  &  &  0 & +20 & 114 & " &  - & - & - & -& -\\  
 & CS$(3-2)$ & 146.969 & 20:34:51.88 & 60:09:14.9 & 0 & 0 & 189 & 16.7 & 1.2 $\pm$ 0.1 & 48.1 $\pm$ 3.9 & 23.0 & 2.1 & 8.0 $\pm$ 1.8 \\ 
 & &  &  &  &  0 & -10 & 271 & " &  - & - & - & -& - \\  
  & &  &  &  &  0 & +10 & 209 & " &   - & - & - & -& - \\ 
 & &  &  &  &  +10 & 0 & 230 & " & - & - & - & -& - \\  
 & &  &  &  &  -10 & 0 & 202 & " &  - & - & - & -& - \\  
 & &  &  &  &  -20 & 0 & 156 & " & 0.4 $\pm$ 0.1 & 46.6 $\pm$ 8.5 & 7.9 & 2.8 & 1.4  $\pm$ 5.1\\
 & &  &  &  &  +20 & 0 & 134 & " &  - & - & - & - & -\\ 
 & &  &  &  &  +10 & +10 & 205 & " & - & - & - & - & -\\   
 & &  &  &  &  +10 & -10 & 196 & " & - & - & - & - & -\\  
 & &  &  &  &  0 & -20 & 166 & " & - & - & - & - & -\\   
 & &  &  &  &  0 & +20 & 161 & " &  - & - & - & - & -\\   
 
 \hline
 NGC 6946-2 & CS$(2-1)$ & 97.980 & 20:34:51.88 & 60:09:14.9 & 0 & 0 & 133 & 25.1 & 2.5 $\pm$ 0.2 & 86.7 $\pm$ 7.8 & 26.6 & 2.1& 101.2 $\pm$ 4.1 \\
 & &  &  &  &  0 & -10 & 132 & " & 1.5 $\pm$ 0.1 & 80.4 $\pm$ 8.2 & 17.8 & 2.1 & 88.0 $\pm$ 4.0 \\     
 & &  &  &  &  0 & +10 & 131 & " &  - & - & - & - & -\\  
 & &  &  &  &  +10 & 0 & 140 & " & 1.6 $\pm$ 0.3 & 73.3 $\pm$ 13.5 & 19.9 & 2.5 & 94.9 $\pm$ 3.2\\ 
 & &  &  &  &  -10 & 0 & 127 & " & 1.4 $\pm$ 0.1 & 77.8 $\pm$ 9.7 & 16.8 & 2.0 & 96.2 $\pm$ 3.5 \\ 
 & &  &  &  &  -20 & 0 & 113 & " &  - & - & - & - & -\\  
 & &  &  &  &  +20 & 0 & 104 & " & 0.5 $\pm$ 0.1 & 83.0 $\pm$ 19.3 & 6.0 & 2.2 & 85.7 $\pm$ 9.9 \\   
 & &  &  &  &  +10 & +10 & 130 & " &  - & - & - & - & -\\  
 & &  &  &  &  +10 & -10 & 125 & " &  - & - & - & - & -\\  
 & &  &  &  &  0 & -20 & 117 & " &  - & - & - & - & -\\  
 & &  &  &  &  0 & +20 & 114 & " &  - & - & - & - & -\\ 

 & CS$(3-2)$ & 146.969 & 20:34:51.88 & 60:09:14.9 & 0 & 0 & 189 & 16.7 & 2.4 $\pm$ 0.1 & 79.2 $\pm$ 5.1 & 28.6 & 2.1 & 87.1 $\pm$ 1.9 \\ 
 & &  &  &  &  0 & -10 & 271 & " &  - & - & - & - & -\\  
  & &  &  &  &  0 & +10 & 209 & " &   - & - & - & - & - \\ 
 & &  &  &  &  +10 & 0 & 230 & " & - & - & - & - & - \\  
 & &  &  &  &  -10 & 0 & 202 & " &  - & - & - & - & - \\  
 & &  &  &  &  -20 & 0 & 156 & " & 0.3 $\pm$ 0.1 & 54.6 $\pm$ 21.4 & 5.6 & 2.8 & 83.4 $\pm$ 7.4 \\
 & &  &  &  &  +20 & 0 & 134 & " &  - & - & - & -& - \\  
 & &  &  &  &  +10 & +10 & 205 & " & - & - & - & -& - \\   
 & &  &  &  &  +10 & -10 & 196 & " & - & - & - & - & -\\  
 & &  &  &  &  0 & -20 & 166 & " & -  & - & - & -& -\\   
 & &  &  &  &  0 & +20 & 161 & "  & - & - & - & -& -\\

\hline
\hline
\end{tabular}

\end{center}

\end{table*}

\begin{table*}[htbp]
\scriptsize
\begin{center}
\caption{NGC 3079 observational parameters and Gaussian fits. For a two component fit, NGC 3079-1 corresponds to the component at lower velocity and NGC 3079-2 the higher velocity component.}
\label{tab:3079obs}

\begin{tabular}{ c r r H H r r H H r r r r r }
Source & Line & $\nu$ & RA (J2000) & Dec (J2000) & Offset & Offset  & T$_{sys}$ & Beam & $\int$(T$_{mb}$d$\nu$) & FWHM & T$_{peak}$ & rms & v$_{peak}$  \\    
 & & (GHz) & (h:m:s) & ( $^{\circ}$ : ' : " )  & RA (") & Dec (") & (K) & Size (") & (K km s$^{-1}$) & (km s$^{-1}$) & (mK) & (mK) & (km s$^{-1}$) \\
\hline\hline
NGC 3079 & CS$(2-1)$ & 97.980 & 10:01:57.79 & 55.40:47.0 & 0 & 0 & 148 & 25.1 & 2.2 $\pm$ 0.2 & 429.5 $\pm$ 34.6 & 4.7 & 1.4 &1153 $\pm$ 15 \\ 
 & &  &  &  &  0 & -10 & 143 & " &   - & - & - & - & - \\ 
 & &  &  &  &  0 & +10 & 151 & " & 2.7 $\pm$ 0.4 & 464  $\pm$ 72 & 5.4 & 2.5 & 1213 $\pm$ 26 \\
 & &  &  &  &  +10 & 0 & 140 & " & 2.4 $\pm$ 0.3 & 472 $\pm$ 67 & 4.8 & 2.6 & 1149 $\pm$ 26 \\
 & &  &  &  &  -10 & 0 & 139 & " &   - & - & - & - & - \\  
 & &  &  &  &  -10 & +10 & 157 & " &   - & - & - & - & - \\ 
 & &  &  &  &  -20 & +20 & 162 & " &   - & - & - & - & - \\ 
 & CS$(3-2)$ & 146.969 & 10:01:57.79 & 55.40:47.0 & 0 & 0 & 234 & 25.1 & 4.3 $\pm$ 0.3 & 484.9 $\pm$ 40.8 & 8.2 & 2.5  & 1140 $\pm$ 14 \\
 
 & H$_{2}$CO(2(0,2)-1(0,1)) & 145.603 &  20:34:51.88 & 60:09:14.9   &  0 & 0 & 189 & 16.7 & 1.9 $\pm$ 0.2 & 444 $\pm$ 59 & 4.0 & 2.7 & 1190 $\pm$ 28 \\
 & CH$_{3}$OH($2_{k}-1_{k}$) & 96.745 &  20:34:51.88 & 60:09:14.9  &  0 & 0 & 133 & 25.1 & 0.9 $\pm$ 0.1 & 362 $\pm$ 61 & 2.3 & 1.1 & 1240 $\pm$ 31 \\
  & CH$_{3}$OH($3_{k}-2_{k}$) & 145.125 &  20:34:51.88 & 60:09:14.9   &  0 & 0 & 189 & 16.7 & 2.0 $\pm$ 0.3 & 394 $\pm$ 64 & 4.7 & 3.2 & 1158 $\pm$ 27  \\ 
  
\hline

NGC 3079-1 & CS$(2-1)$ & 97.980 & 10:01:57.79 & 55.40:47.0 & 0 & 0 & 148 & 25.1 & 0.9 $\pm$ 0.2 & 208.3 $\pm$ 42.1 & 4.3 & 1.5 & 1010 $\pm$ 21 \\ 
 & CS$(3-2)$ & 146.969 & 10:01:57.79 & 55.40:47.0 & 0 & 0 & 234 & 25.1 & 2.0 $\pm$ 0.4 & 244.4 $\pm$ 41.1 & 7.8 & 2.5 & 994 $\pm$ 22 \\

\hline
NGC 3079-2 & CS$(2-1)$ & 97.980 & 10:01:57.79 & 55.40:47.0 & 0 & 0 & 148 & 25.1 & 1.2 $\pm$ 0.2 & 222.0 $\pm$ 35.4 & 4.8 & 1.5 & 1270 $\pm$ 20  \\ 
& CS$(3-2)$ & 146.969 & 10:01:57.79 & 55.40:47.0 & 0 & 0 & 234 & 25.1 & 2.1 $\pm$ 0.4 & 230.3 $\pm$ 37.6 & 8.5 & 2.5 &1279 $\pm$ 19 \\
\hline

\end{tabular}

\end{center}
\end{table*}

\section{Results and analysis}
\subsection{Results}

\begin{figure*}[htbp]
\begin{center}
\includegraphics[width=0.25\linewidth,angle=270]{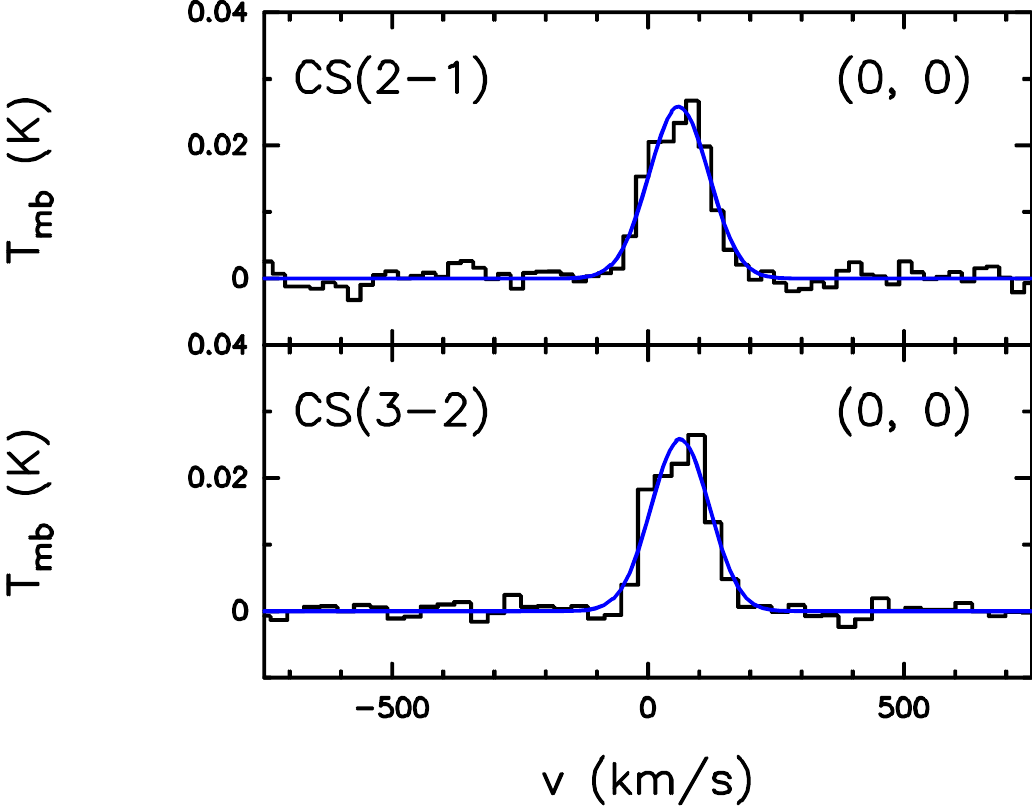}
\includegraphics[width=0.25\linewidth,angle=270]{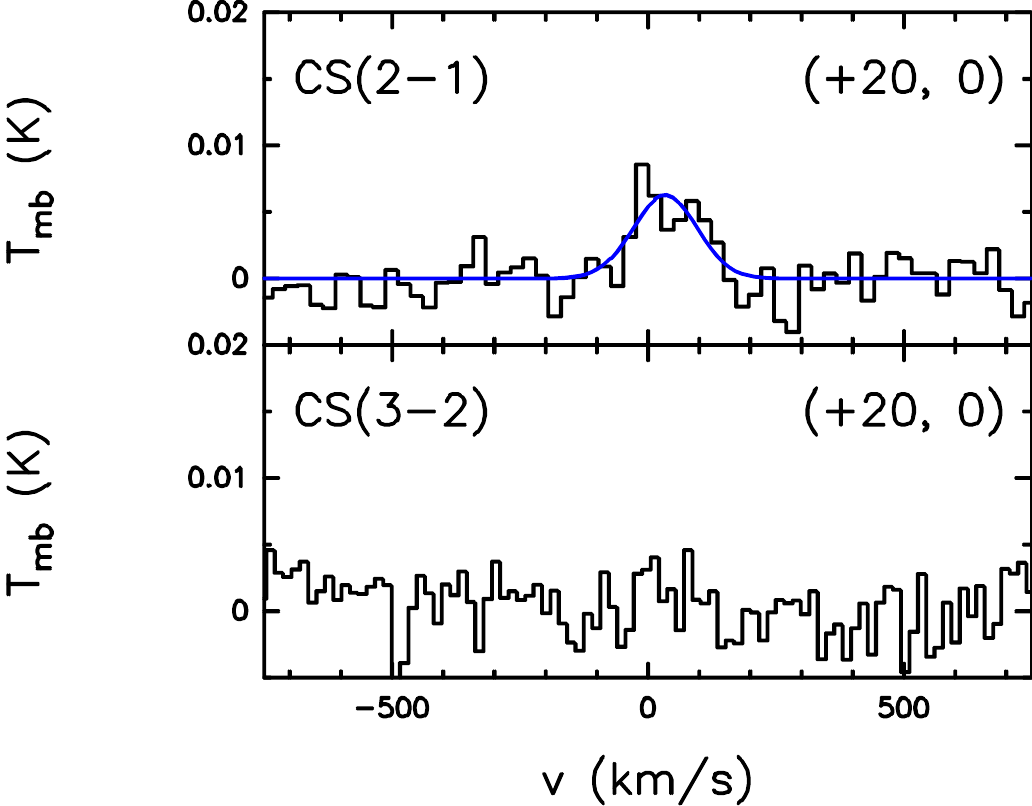}
\includegraphics[width=0.25\linewidth,angle=270]{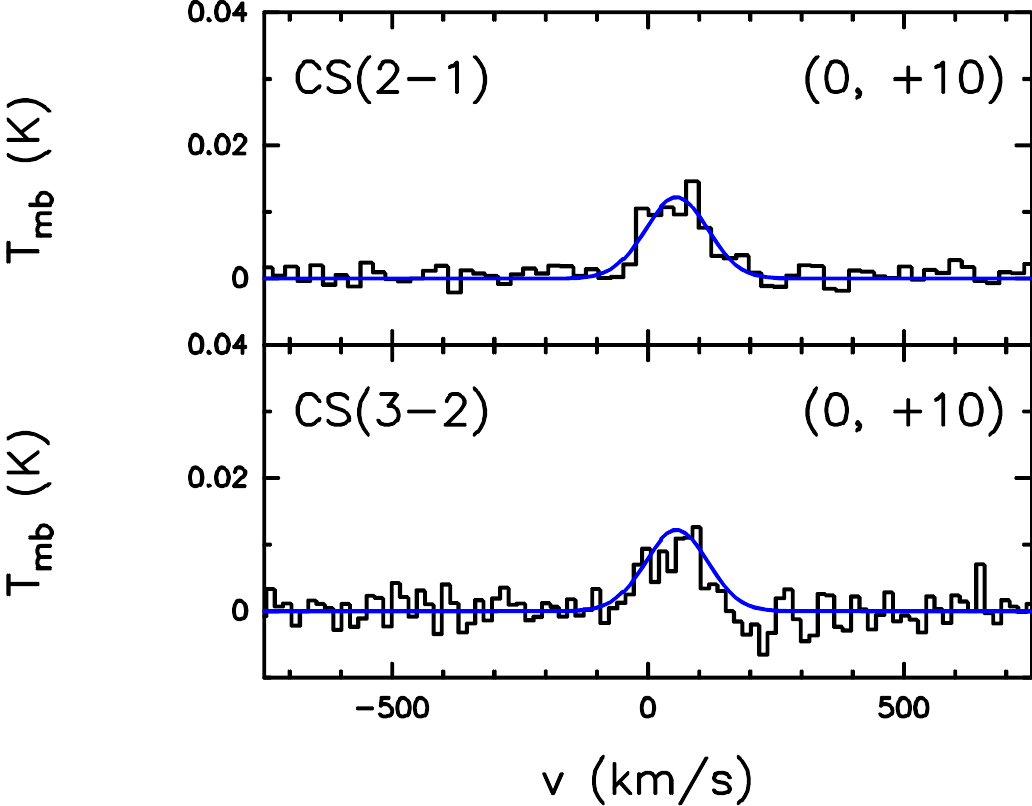}
\includegraphics[width=0.25\linewidth,angle=270]{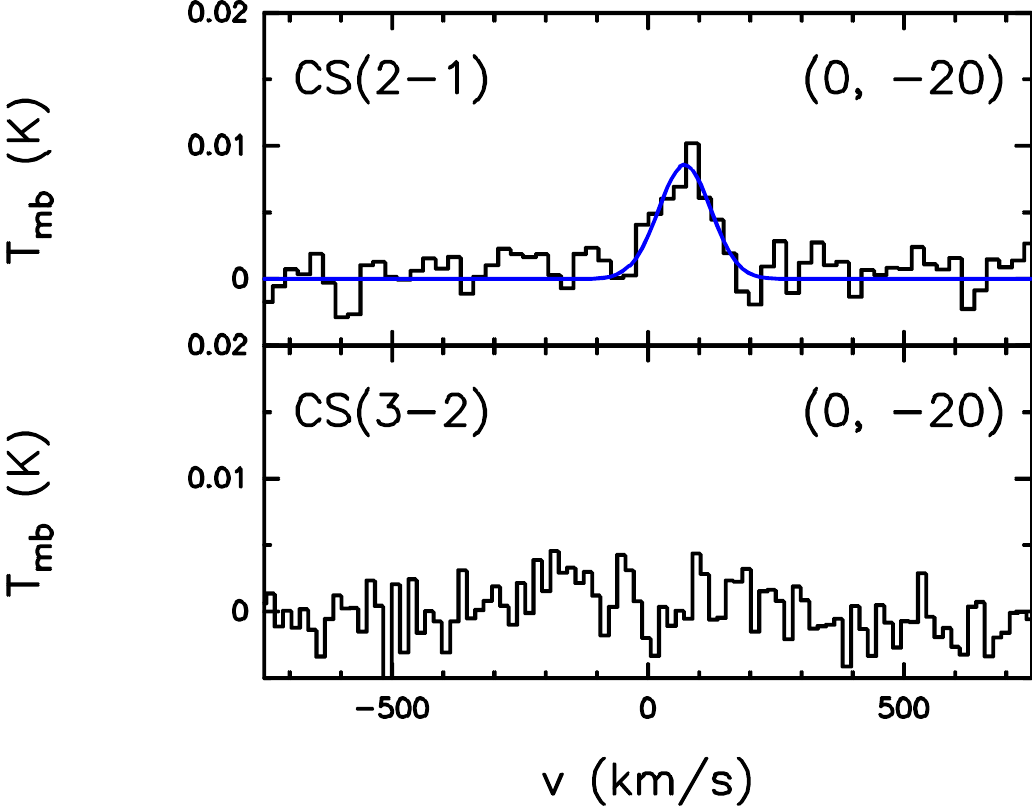}
\includegraphics[width=0.25\linewidth,angle=270]{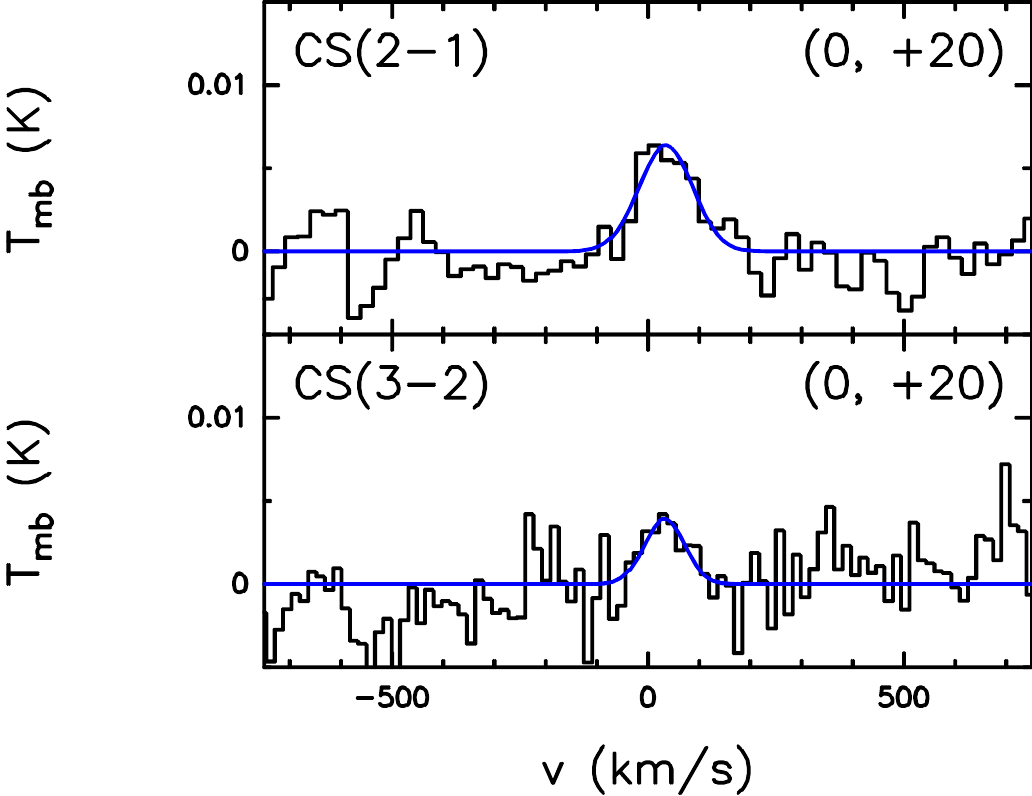}
\includegraphics[width=0.25\linewidth,angle=270]{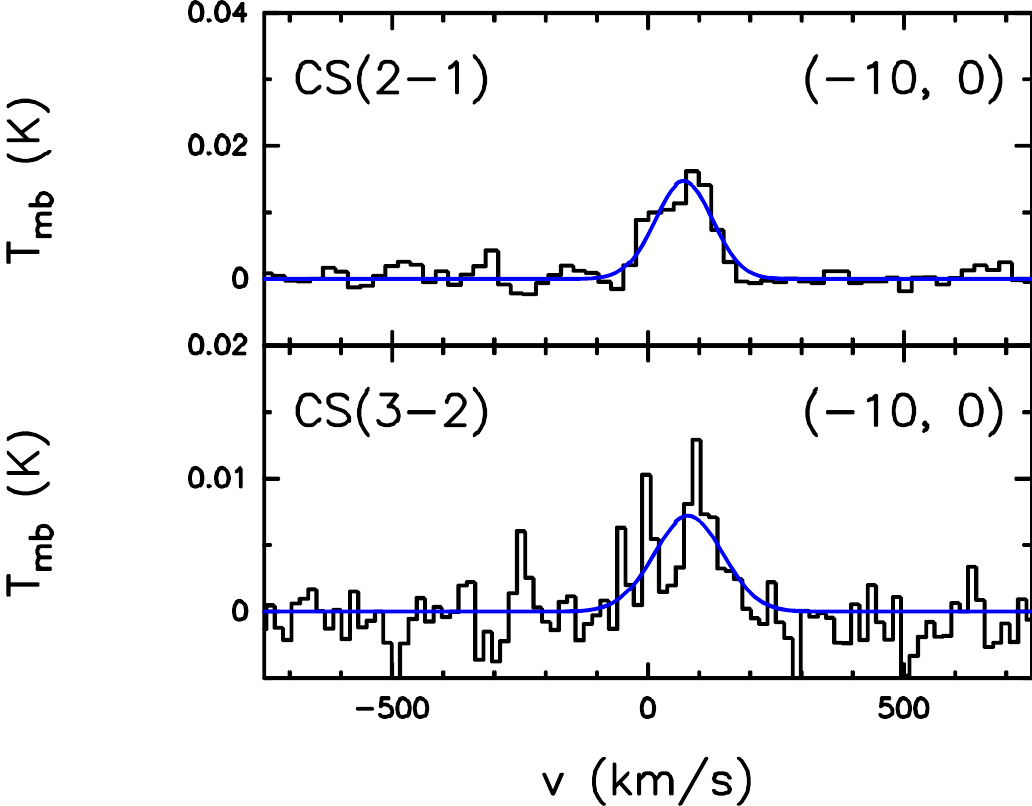}
\includegraphics[width=0.25\linewidth,angle=270]{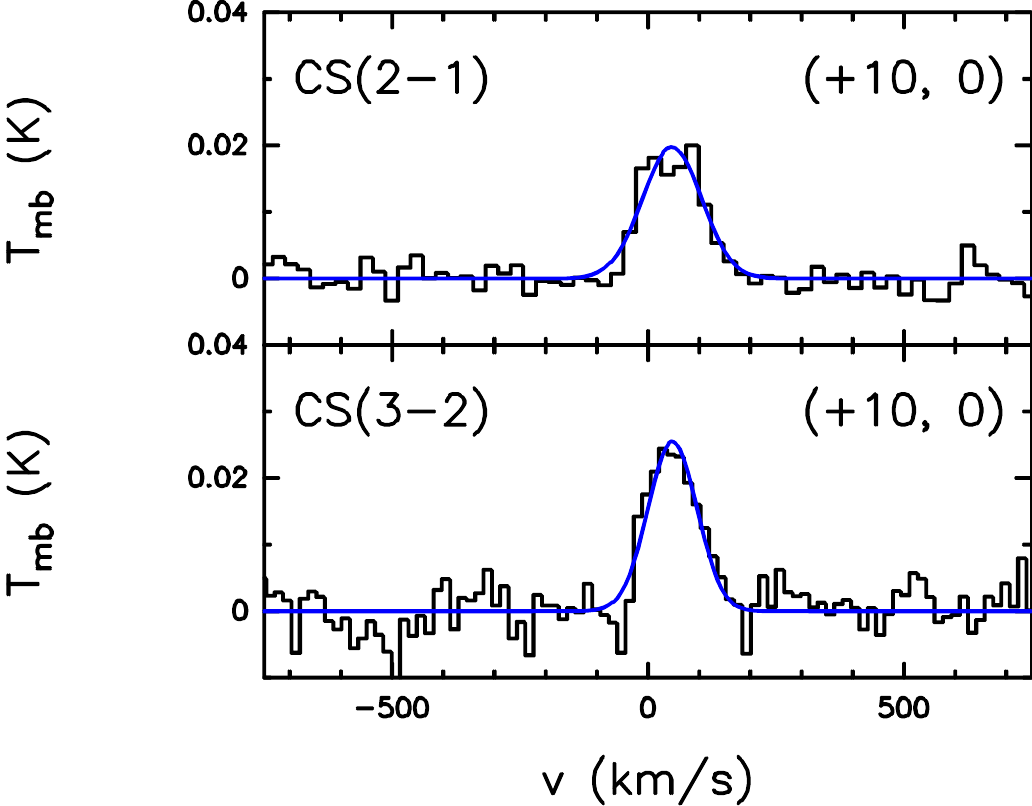}
\includegraphics[width=0.25\linewidth,angle=270]{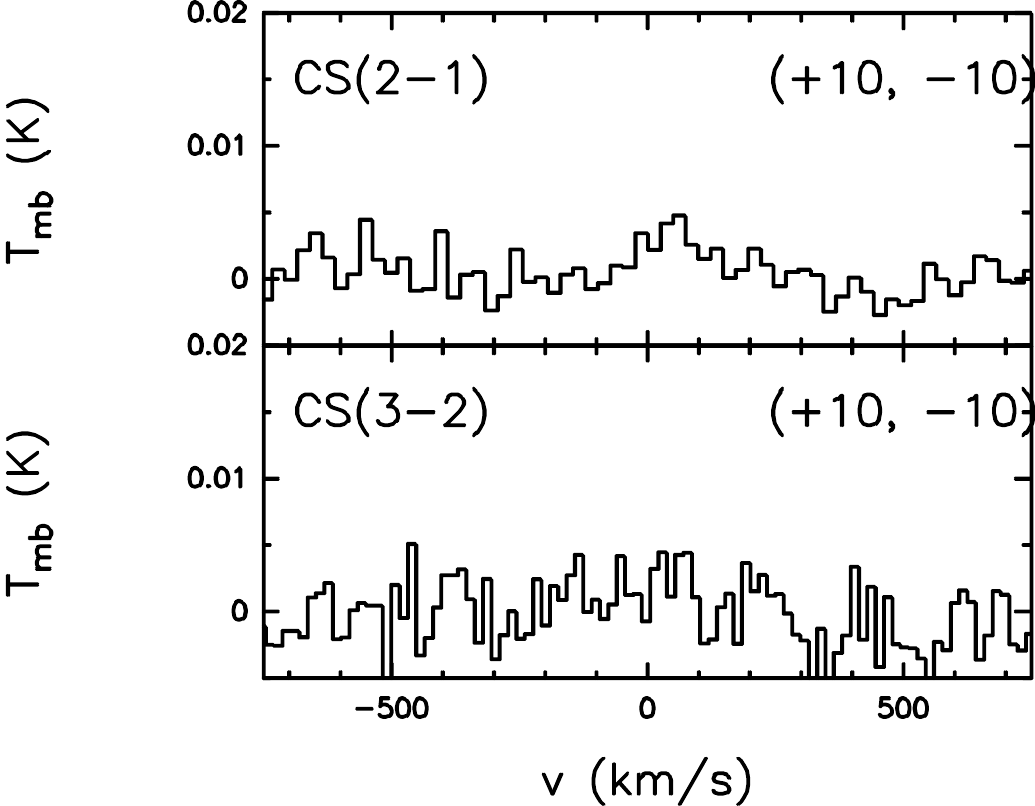}
\includegraphics[width=0.25\linewidth,angle=270]{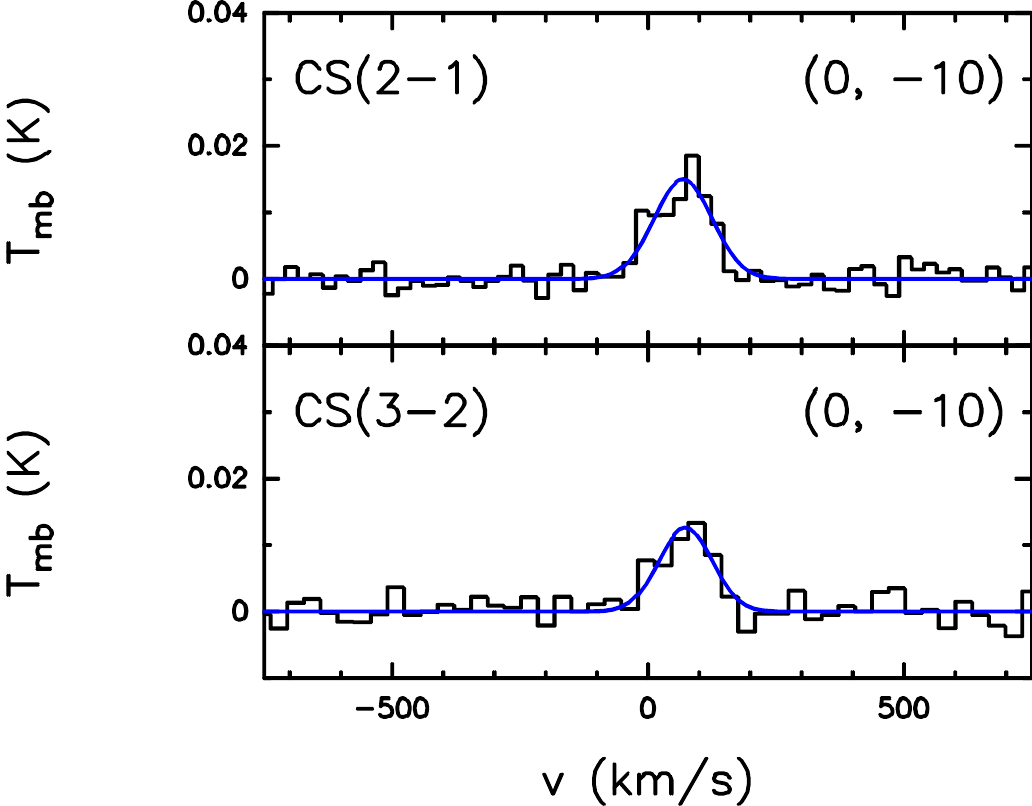}
\includegraphics[width=0.25\linewidth,angle=270]{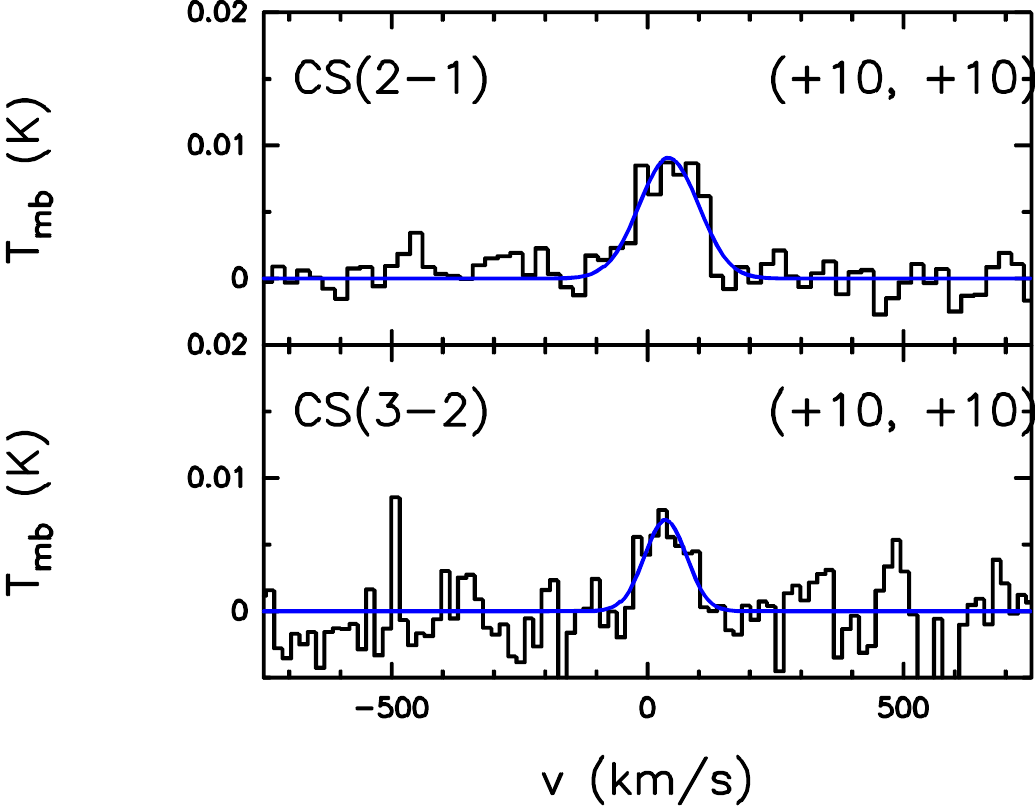}
\includegraphics[width=0.25\linewidth,angle=270]{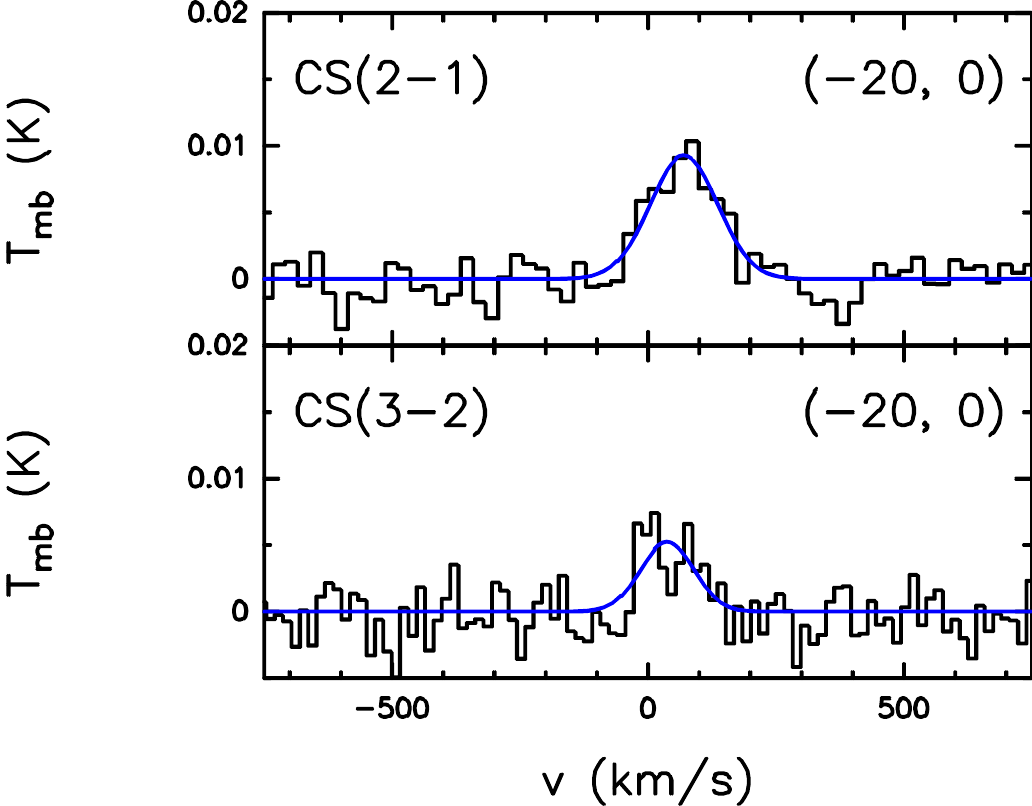}

\caption{Single component fits for CS$(2-1)$ and CS$(3-2)$ in NGC\,6946. Label in right corner shows offset in RA ('') and Dec ('') from central position.}
\label{fig:csdetections6}
\end{center}
\end{figure*}

\begin{figure*}[htbp]
\begin{center}
\includegraphics[width=0.25\linewidth,angle=270]{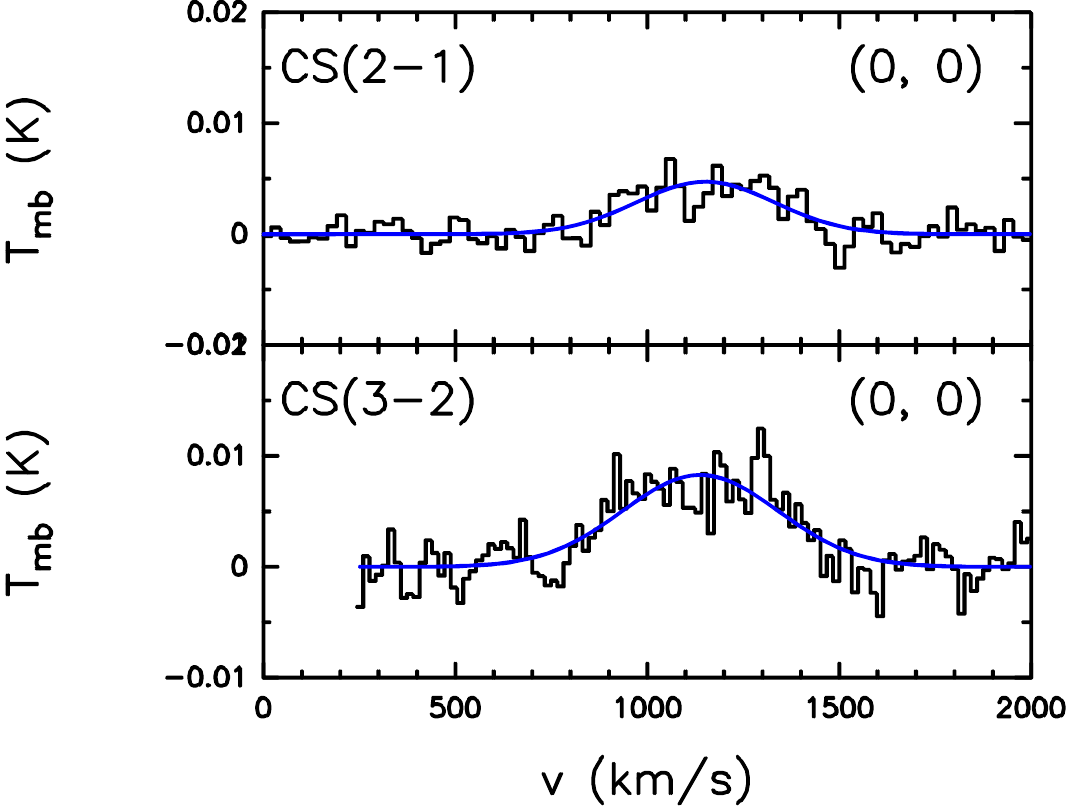}
\includegraphics[width=0.25\linewidth,angle=270]{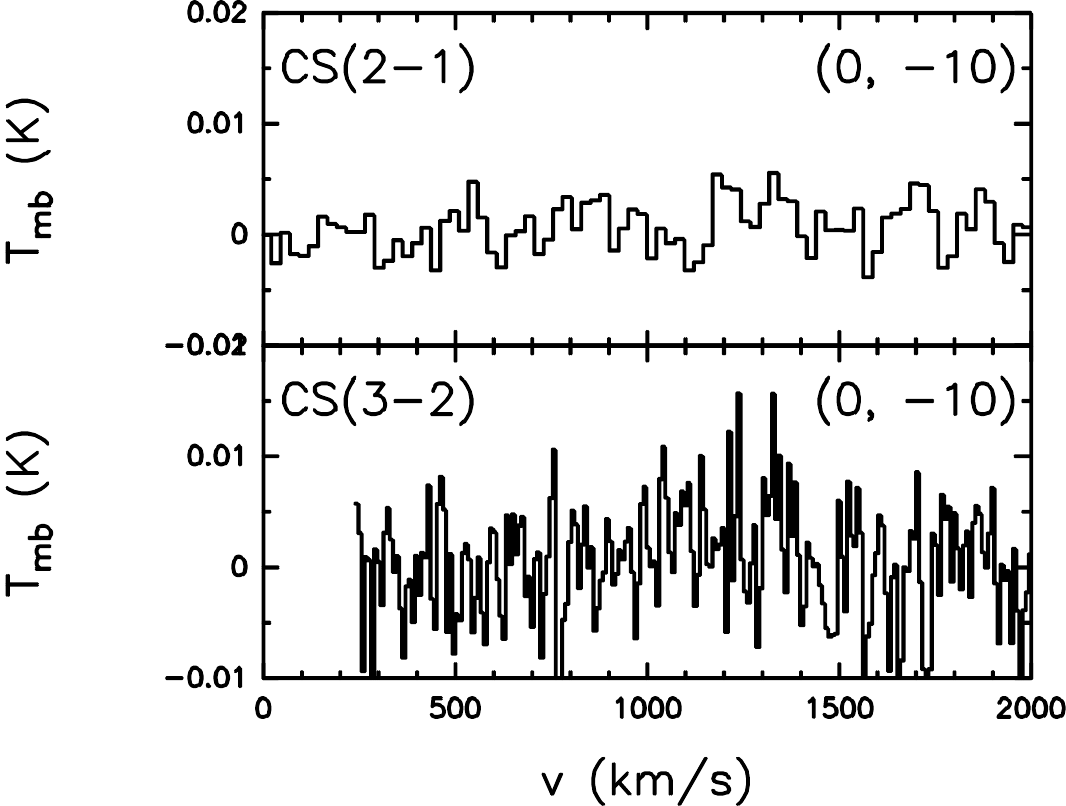}
\includegraphics[width=0.25\linewidth,angle=270]{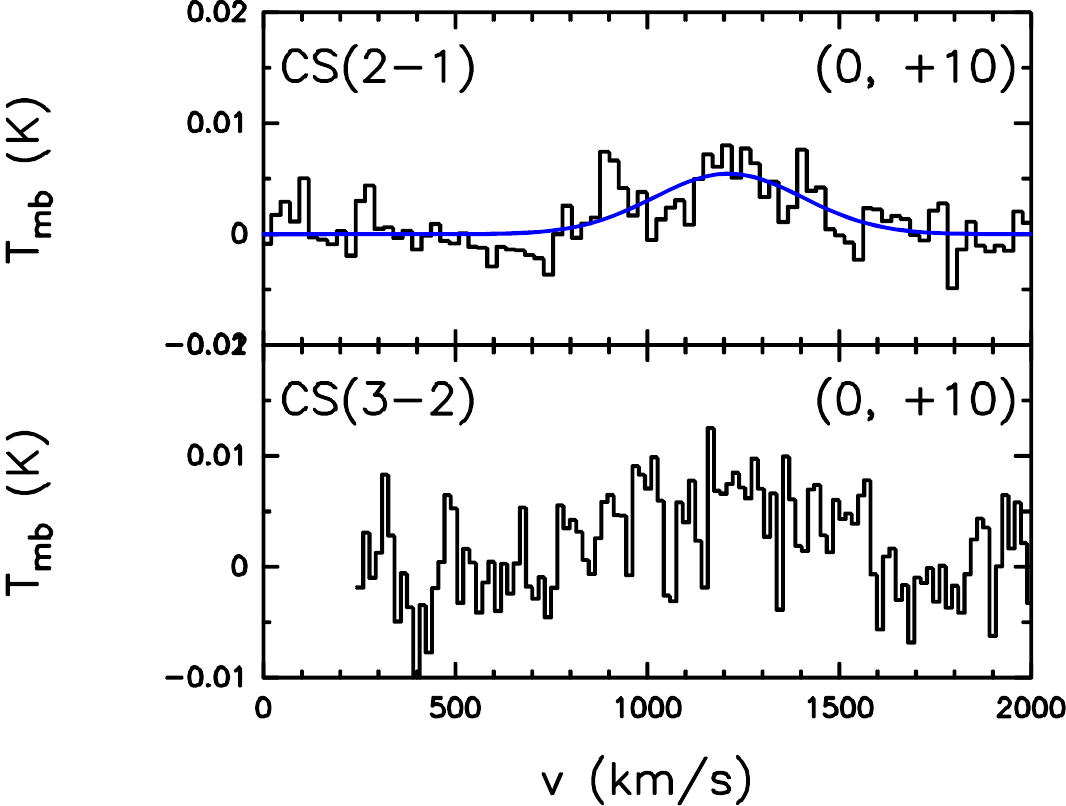}
\includegraphics[width=0.25\linewidth,angle=270]{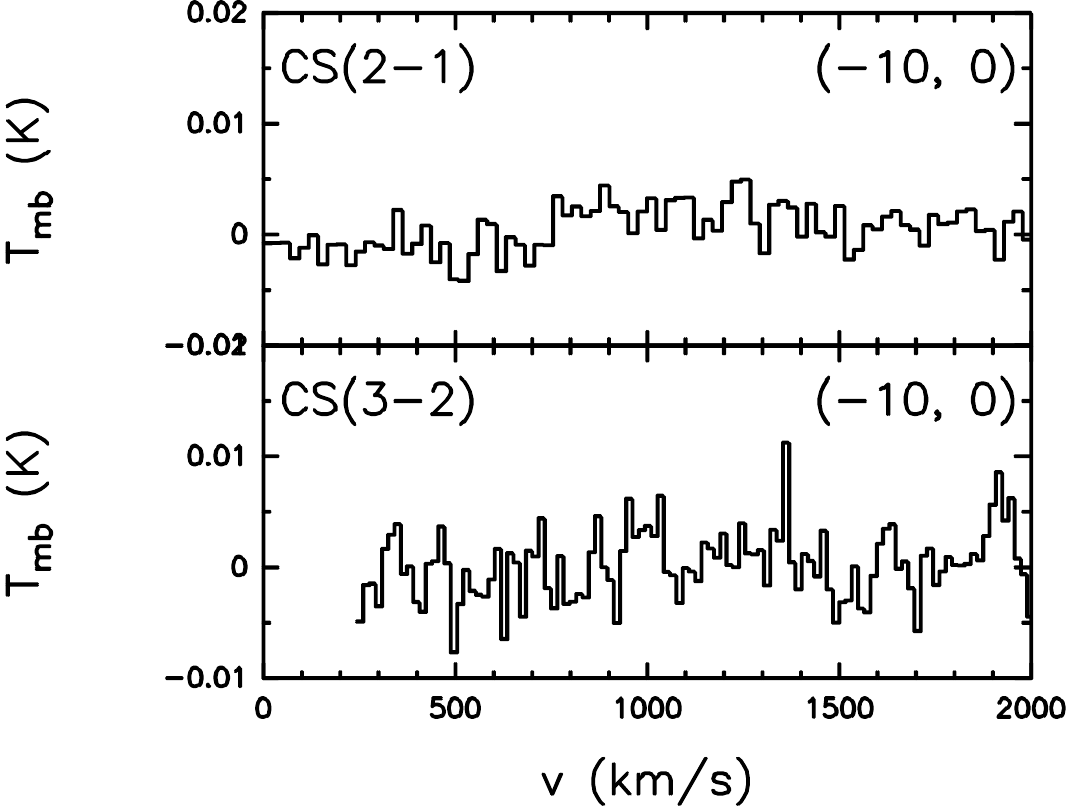}
\includegraphics[width=0.25\linewidth,angle=270]{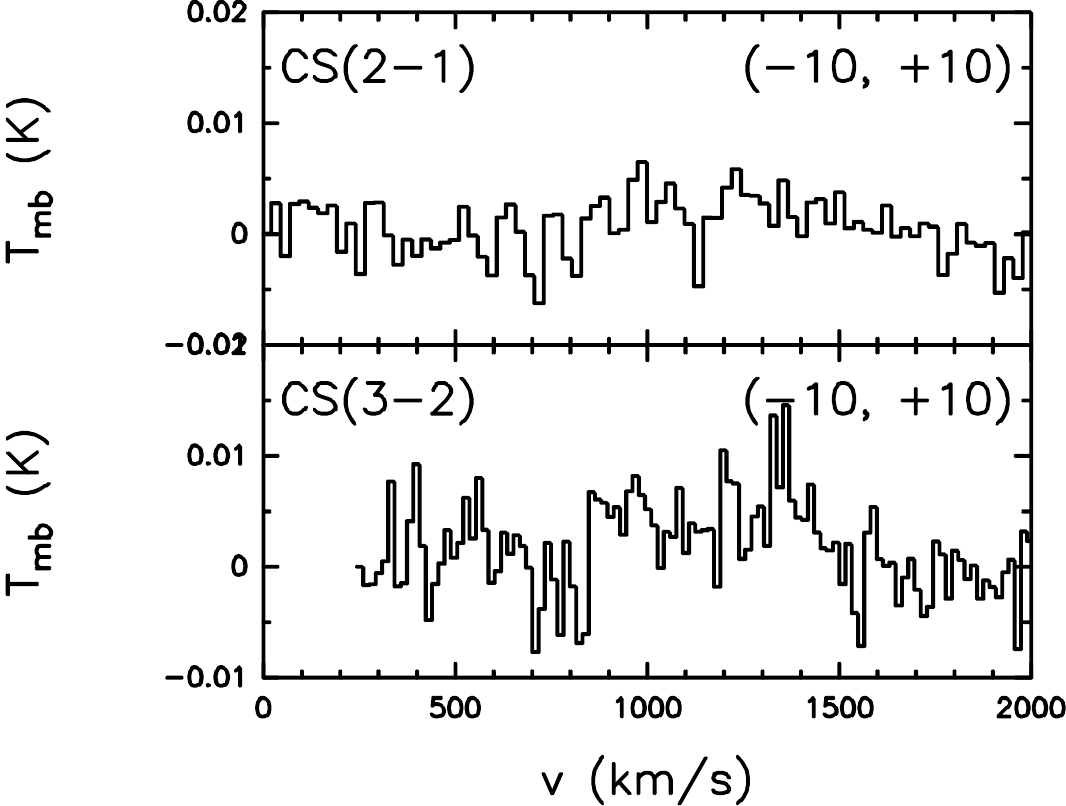}
\includegraphics[width=0.25\linewidth,angle=270]{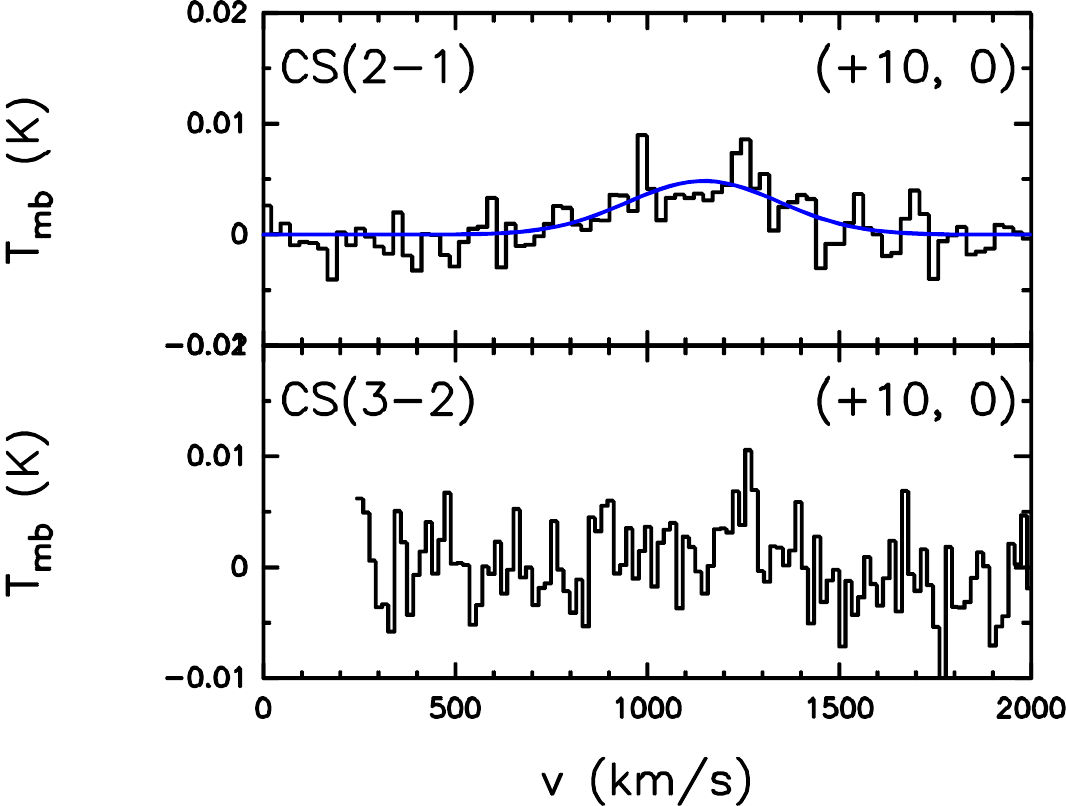}
\includegraphics[width=0.25\linewidth,angle=270]{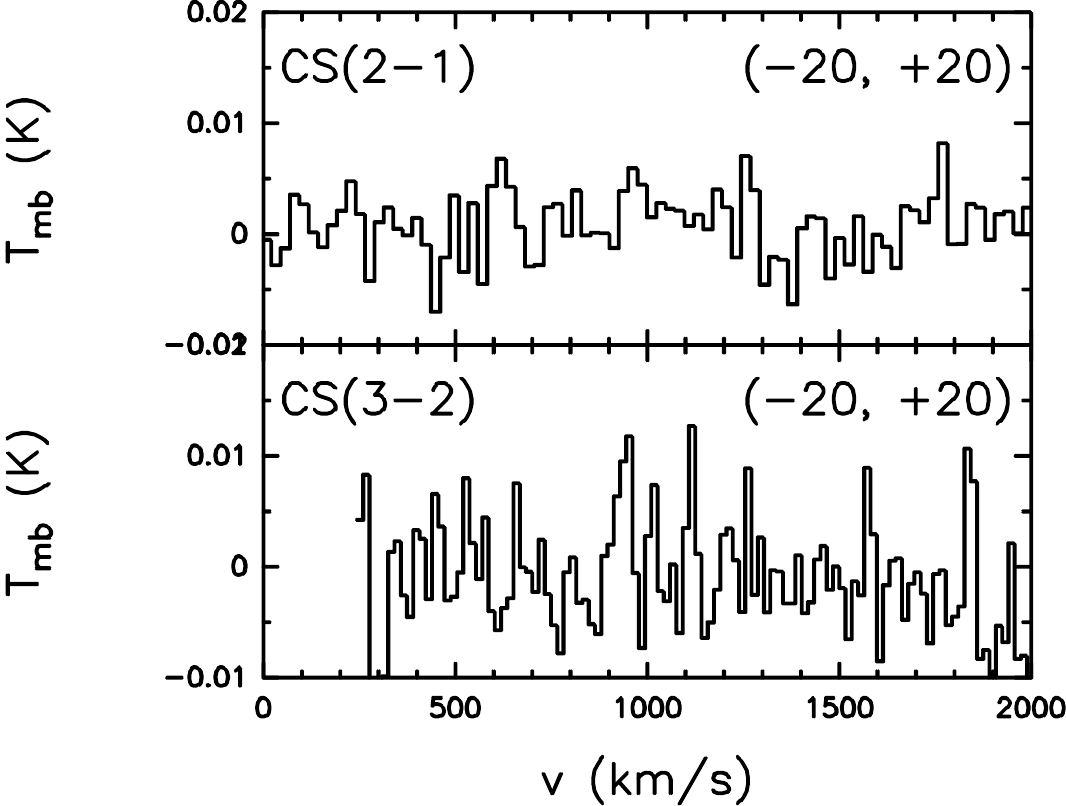}
\caption{Single component fits for CS$(2-1)$ and CS$(3-2)$ in NGC\,3079. Label in right corner shows offset in RA ('') and Dec ('') from central position.}
\label{fig:csdetections3}
\end{center}
\end{figure*}

Maps of CS spectra for both galaxies are shown in Figure ~\ref{fig:maps}. Broadly, the strongest emission is seen in the central position (see Tables ~\ref{tab:6946obs} and ~\ref{tab:3079obs} for coordinates) of both galaxies, with two exceptions. In NGC 3079, the CS$(2-1)$ line is stronger when the beam is offset by +10'' in Declination (Dec), and separately, when offset by +10'' in Right Ascension (RA). NGC 6946 shows the strongest emission in the centre, getting weaker out towards offsets of $\pm$20'' in all directions. \citet{1989A&A...226L...5M} discovered CS in NGC 6946 and found CS$(3-2)$ in offset position (0, 0) and (10, -10) from coordinates $\alpha_{2000}$ = 20:34:51 and $\delta_{2000}$ = 60:09:25. This is a few arc seconds from our central position. Our results appear to agree with this but ours also show strong CS emission from other regions, most notably from the offset of +10'' in RA. We fit single Gaussian profiles to all detections of CS (Figure ~\ref{fig:csdetections6} \& ~\ref{fig:csdetections3}). We also detect methanol (CH$_{3}$OH) and formaldehyde (H$_{2}$CO) in both galaxies. The isotoplogue of CS, C$^{34}$S was marginally detected, also in both galaxies. Tables ~\ref{tab:6946obs} and ~\ref{tab:3079obs} show the Gaussian fit parameters for these detections.

Since we have no prior knowledge as to the location of CS emission, we consider each position as separate and assume the source fills the beam. The signal-to-noise (S/N) of CS in some offset positions is less than 3, and therefore they are considered as tentative detections. Still, we used them in our analysis, as the fitted line widths and velocity positions agree with the values expected. NGC 6946 and NGC 3079 angular sizes are $\approx$ 12' x 8' and $\approx$ 4.5' x 1' in the IR \citep{2006AJ....131.1163S} so observations focus on the area surrounding the centre. It is interesting to note that in offset positions of just $\pm$20'' in any direction in both galaxies we fail to see (or only marginally detect) CS. \citet{1985ApJ...298L..21B} found a bar structure, running approximately north-south in NGC 6946 of size 11'' by 55''. More material in this bar may mean that there is more dense gas, so it might be expected that some CS is detected further away from the centre, in at least one direction. 

\subsection{Line analysis}

Under the assumption that emission from different molecules is coming from the same regions, we calculate velocity integrated line ratios for some species. We note that this assumption may not hold completely true, but include the values as they may be instructive. We calculate the ratio of CS$(3-2)$ / CS$(2-1)$ where possible, accounting for the difference in beam sizes between the two detections. Since we only record detections of lines other than CS$(2-1)$ in our central pointing in NGC 3079, we only find ratios for the central position in this galaxy. These are listed in Table ~\ref{tab:otherratios}, together with ratios in the central position of NGC 6946.  A map of CS$(3-2)$ / CS$(2-1)$ ratios in NGC 6946 is shown in Table ~\ref{tab:csratios}. In NGC 3079 we see much stronger emission from CS$(3-2)$ when compared to all other lines. In the central position, the CS$(3-2)$ / CS$(2-1)$ ratio is approximately double the equivalent in NGC 6946. The same is seen when comparing CS$(3-2)$ to other lines. In NGC 6946 the CS$(3-2)$ / CS$(2-1)$ ratio is highest at offset +10'' in RA (approximately 270 pc). Other than this, the general trend is for these values to drop as the beam is pointed away from the galactic centre, although the value is always \textgreater 1.

\begin{figure}[htbp]
\begin{center}
\includegraphics[width=1.1\linewidth,angle=270]{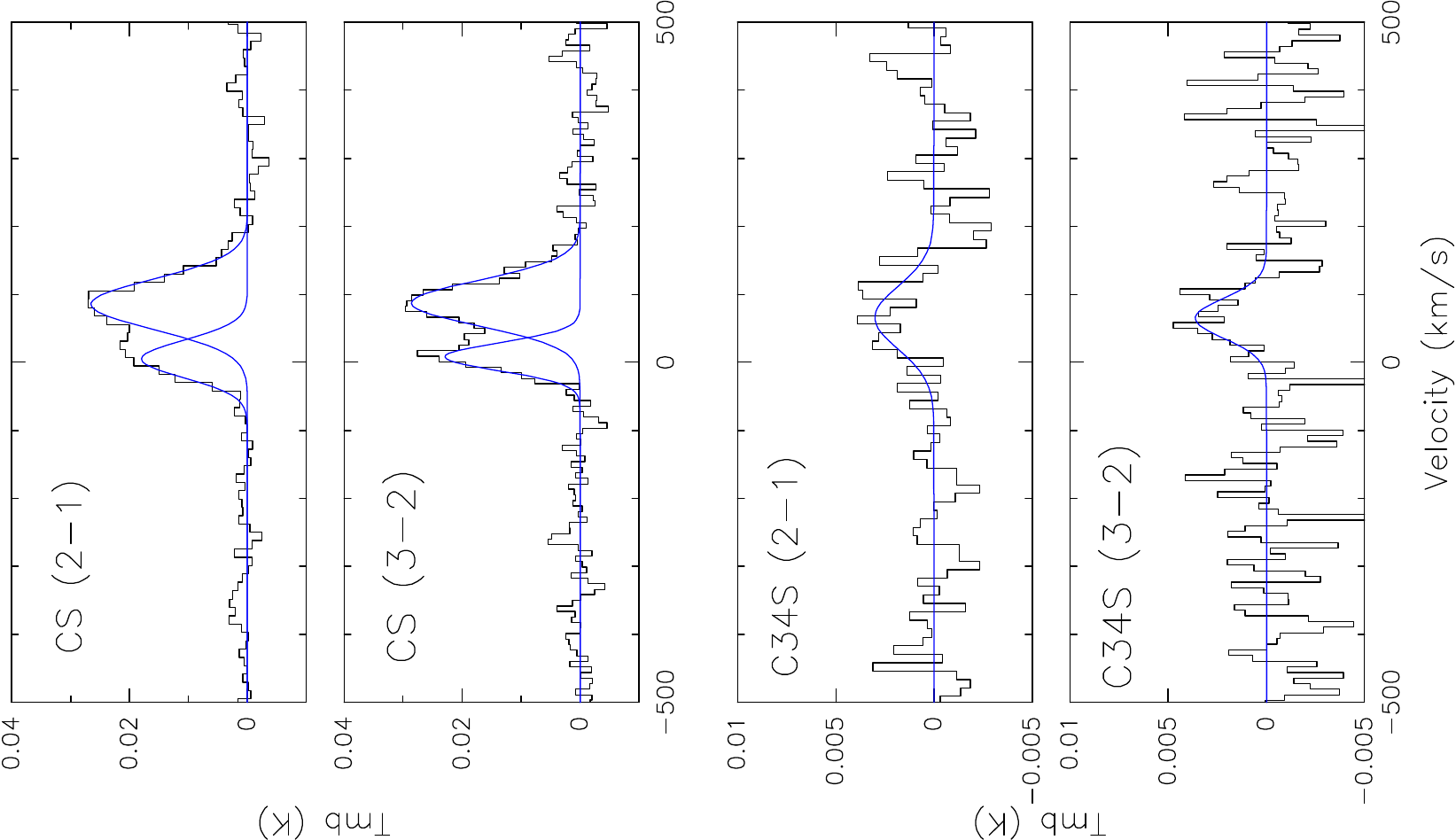}
\caption{CS detections in centre of NGC 6946. $^{12}C^{32}S$ displaying a double-horn profile.}
\label{fig:csdoublehorn6946}
\end{center}
\end{figure}

\begin{figure}[htbp]
\begin{center}
\includegraphics[width=0.55\linewidth,angle=270]{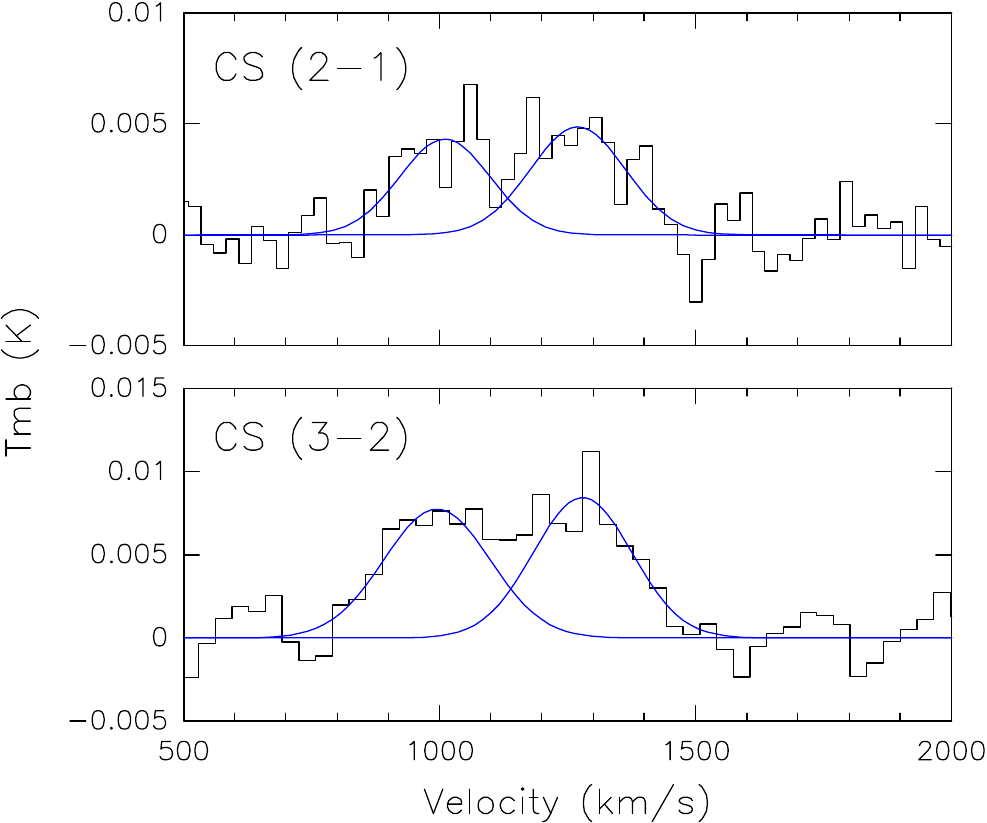}
\caption{CS detections in centre of NGC 3079. $^{12}C^{32}S$ displaying a double-horn profile.}
\label{fig:csdoublehorn3079}
\end{center}
\end{figure}

Although we have treated lines as being a single peak, when the data is less smoothed, detected CS lines in both galaxies show a double peak profile (Figure ~\ref{fig:csdoublehorn6946} \& ~\ref{fig:csdoublehorn3079}), although this is not very clear in the tentative detections. We consider a possible cause of this to be a rotating torus of dense gas around the nucleus, leading to red- and blue-shifted components. Profiles like these have been observed in NGC 253 \citep{2005ApJ...620..210M} leading the authors to a similar conclusion. However, we note that this might be due to other factors, such as self absorption. The lower velocity (red-shifted) component in both galaxies is narrower, possibly meaning that the gas moving away has a lower turbulent velocity than the gas coming towards. We show the double-horn Gaussian fits for the centre of each galaxy and compare with the isotopolgue C$^{34}$S in NGC 6946 (Figure ~\ref{fig:csdoublehorn6946}). Detections of isotopologues were marginal (\textless 2 $\sigma$) in NGC 3079 so only the double-horn CS profile is fitted (Figure  ~\ref{fig:csdoublehorn3079}). These data have been included with the single peak data in Table ~\ref{tab:6946obs} and ~\ref{tab:3079obs}. For both single and double peak profiles, a good agreement can be found between the Gaussian fits and the observations. Where possible we fit to the double peak but we also calculate for a single fit. If a double fit is not possible (as with many of the offset points), we only fit a single Gaussian. For double fits we have designated the component at lower velocity 1 and the higher velocity component 2 (e.g. NGC 6946-1 and NGC 6946-2).

\begin{table}[htbp]
\begin{center}
\caption{Other line ratios for central pointing}
\label{tab:otherratios}
\begin{tabular}{c c c}
Galaxy & Lines & Ratio \\
\hline
 NGC 6946 & {CS$(3-2)$ / CS$(2-1)$} & {3.2 $\pm$ 0.2} \\
 & CH$_{3}$OH($2_{k}-1_{k}$) / CS$(2-1)$ & 0.4 $\pm$ 0.1\\
 & CH$_{3}$OH($3_{k}-2_{k}$) / CS$(3-2)$ & 0.6 $\pm$ 0.1 \\
 & CH$_{3}$OH($2_{k}-1_{k}$) / CS$(3-2)$ & 0.1 $\pm$ 0.006 \\
 & CH$_{3}$OH($3_{k}-2_{k}$) / CS$(2-1)$ & 2.0 $\pm$ 0.1 \\
\hline
NGC 3079 & CS$(3-2)$ / CS$(2-1)$ & 6.7 $\pm$ 0.3 \\
& CH$_{3}$OH($2_{k}-1_{k}$) / CS$(2-1)$ & 0.4 $\pm$ 0.1\\
& CH$_{3}$OH($3_{k}-2_{k}$) / CS$(3-2)$ & 0.5 $\pm$ 0.1 \\
 & CH$_{3}$OH($2_{k}-1_{k}$) / CS$(3-2)$ & 0.06 $\pm$ 0.003 \\
 & CH$_{3}$OH($3_{k}-2_{k}$) / CS$(2-1)$ & 3.0 $\pm$ 0.2 \\
\end{tabular}
\end{center}
\end{table}

\subsection{LTE analysis}
\begin{figure}[htbp]
\begin{center}
\includegraphics[width=0.6\linewidth,angle=270]{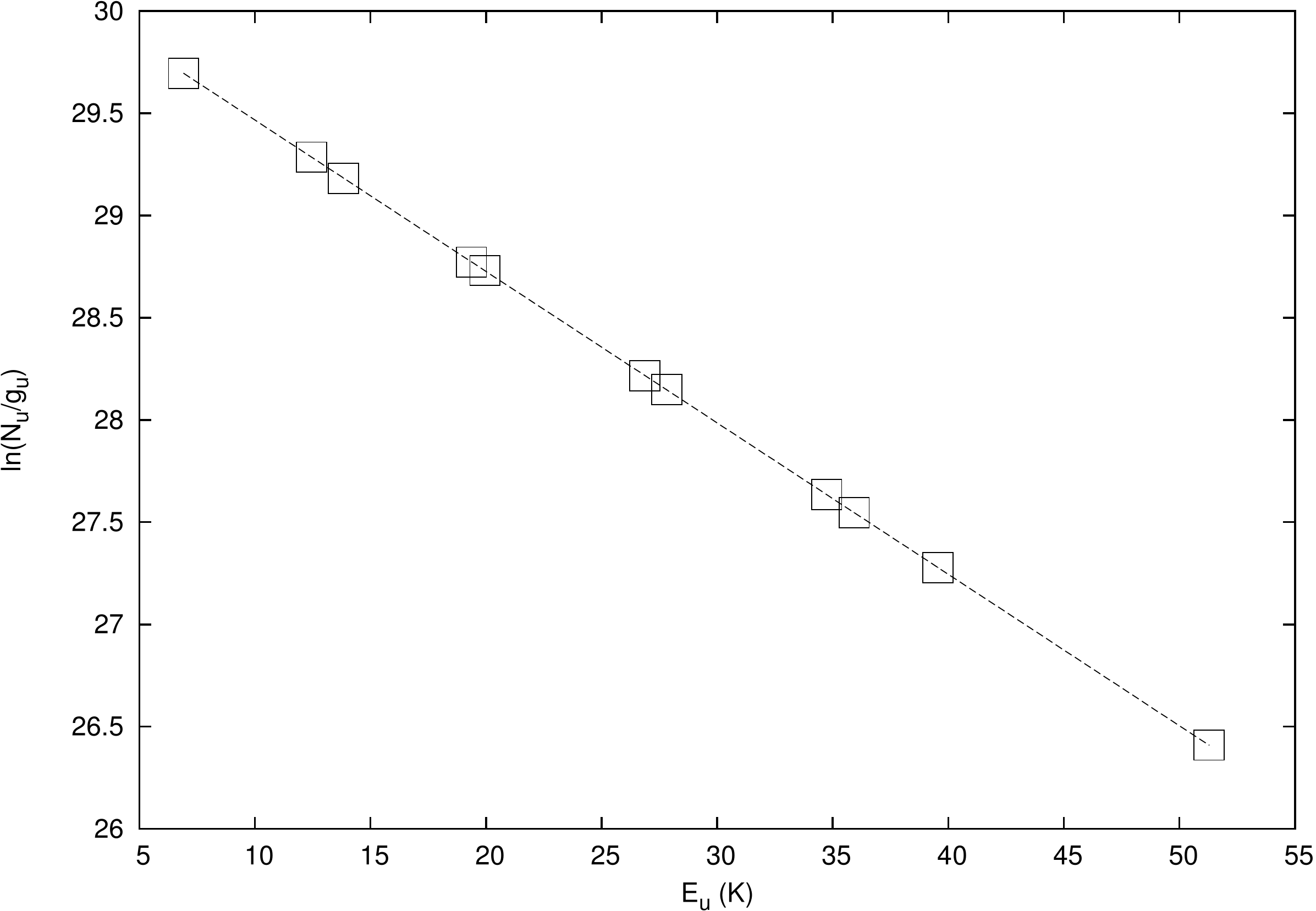}
\caption{{Rotation diagram for methanol in centre of NGC 6946. The parameters for this diagram are T = 13.6 K and N(CH$_{3}$OH) = \num{4.6e14} cm$^{-2}$.}}
\label{fig:methrot}
\end{center}
\end{figure}

We used two methods to perform an LTE analysis of our observations. Firstly, we used the rotational diagram method \citep{1999ApJ...517..209G}. With a critical density for low-J transitions at low temperature of $\approx$ 10$^{5}$ cm$^{-3}$ methanol is also a high density gas tracer, therefore we use this method to constrain a temperature with which to begin an LTE analysis using CS (n$_{crit} \approx$ \num{1e6} for low-J transitions at low temperature). We do this only for the centre of NGC 6946, as lines here are strong enough to allow the use of this technique. This method assumes the gas is in LTE and is optically thin. If this is the case then the data are well fitted by a linear regression. Under LTE conditions, in the rotational diagram method, we are assuming that the rotational temperature is equal to the excitation of all the observed transitions. {Since the gas may not be thermalised the kinetic temperature may be greater than the rotational temperature \citep{1999ApJ...517..209G}}. A value for the column density can also be obtained using this method. We detected two bands of methanol in our spectra (J = 2$_{k}$ - 1$_{k}$, 3$_{k}$ - 2$_{k}$). \citet{2006ApJS..164..450M}  used a method to separate individual transitions in these bands using the ratios of Einstein coefficients $A'_{ul}$ and $A''_{ul}$, upper energy levels $E'_{u}$ and $E''_{u}$ and degeneracies $g'_{u}$ and $g''_{u}$. This is shown in the equation

\begin{equation}
\frac{W'}{W''} = \frac{\nu''^{2}A'_{ul}g'_{u}}{\nu'^{2}A''_{ul}g''_{u}}e^{(E''_{u}-E'_{u})/kT_{rot}^{0}},
\end{equation}
where W is the integrated flux = $\int T_{\rm mb}$d$\nu$, $\nu$ is the frequency and $T_{rot}^{0}$ is the estimate for the rotational temperature. The method works on the basis that in LTE we should see a linear regression in the rotation diagram, the correct temperature is the one that produces the best fit. We therefore iterate over many temperatures using both methanol bands until we find a best fit. The methanol rotation diagram for the centre of NGC 6946 (Figure ~\ref{fig:methrot}) shows a fit to a temperature of 14 K.

In order to obtain an estimation of the CS column density, {we use only our CS$(2-1)$ detections, as we do not detect CS$(3-2)$ in as many locations as CS$(2-1)$. This allows us to calculate values in locations where we have only one detection. In the same way as with the rotational diagram method, we assume LTE conditions and optically thin emission. However,} it does not constrain a temperature. {A rotational temperature must therefore be assumed.} Since we obtained a lower limit to the kinetic temperature of 14K using methanol, column densities are calculated for 15 K, 50 K and 300 K. These calculations have been carried out using both single- and (where possible) double-fits. They are listed in Table ~\ref{tab:lte}. These results are useful as a guide to our chemical modelling as they represent a lower limit to the column density of CS due to the lack of information about the optical depth. We can therefore discard models which show a column density of CS of less than these values at a particular temperature. These results show that reasonable values of column density (N(CS) $\approx$10$^{14}$ cm$^{-2}$) are seen up to temperatures of 300 K.

\begin{table}[htbp]
\begin{center}
\caption{CS$(3-2)$/CS$(2-1)$ ratios in NGC 6946}
\label{tab:csratios}
\begin{tabular}{c c | c c c c c c}
\renewcommand{\arraystretch}{}
& \multicolumn{7}{c}{RA offset ('')} \\
\multirow{12}{*}{\rotatebox{90}{Dec offset ('')}}& & 20 & 10 & 0 & -10 & -20 \\
\hline
& & \multicolumn{6}{c}{} \\
& 20 & & & 1.5 & & & \\
& & \multicolumn{6}{c}{} \\
& 10 &  & 1.7 & 2.4 & & & \\
& & \multicolumn{6}{c}{} \\
& 0 & & 3.5 & 3.2 & 1.9 & 1.5 & \\
& & \multicolumn{6}{c}{} \\
& -10 & &  & 2.5  & & & \\
& & \multicolumn{6}{c}{} \\
& -20 & & &  & & & \\
\end{tabular}
\end{center}
\end{table}

\begin{table*}[htbp]
\begin{center}

\caption{LTE Column Densities}
\begin{tabular}{c c r r c c c }
\hline
Source & Line & Offset & Offset & \multicolumn{3}{c}{N(CS) x 10$^{14}$}  \\
& & RA (") & Dec (") & T = 15 K & T = 50 K & T = 300 K \\
\hline
NGC 6946 & CS$(2-1)$ & 0&0 & 0.176 & 0.42 & 2.25 \\
& & 0 & -10 & 0.099 & 0.24 & 1.27 \\
 
& & 0 & +10 & 0.087 & 0.21 & 1.11 \\
 
& &+10 & 0 & 0.133 & 0.32 & 1.71 \\
 
& & -10 & 0 & 0.096 & 0.23 & 1.22 \\
 
& & -20 & 0 & 0.071 & 0.17 & 0.91 \\
 
& & +20 & 0  & 0.046 & 0.11 & 0.58 \\
 
& & +10 & +10 & 0.062 & 0.15 & 0.79 \\
 
& & +10 & -10 & - & - & - \\
 
& & 0 & -20 & 0.051 & 0.12 & 0.66 \\
 
& & 0 & +20 & 0.039 & 0.09 & 0.5 \\
\hline
NGC 6946-1 & CS$(2-1)$ & 0 & 0 & 0.057 & 0.14 & 0.73 \\
& &0 &-10 & 0.027 & 0.07 & 0.35 \\
& & 0& +10& - & - & - \\
& & +10&0 & 0.056 & 0.14 & 0.72 \\
& & -10&0 & 0.031 & 0.07 & 0.4 \\
& & -20&0 & - & - & - \\
& & +20& 0& 0.02 & 0.05 & 0.26 \\
& & +10&+10 & - & - & - \\
& &+10 &-10 & - & - & - \\
& &0 &-20 & - & - & - \\
& & 0&+20 & - & - & - \\
\hline
NGC 6946-2 & CS$(2-1)$ & 0 & 0 & 0.111 & 0.27 & 1.43 \\
& &0 &-10 & 0.069 & 0.17 & 0.89 \\
& & 0& +10& - & - & - \\
& & +10&0 & 0.071 & 0.17 & 0.9 \\
& & -10& 0& 0.063 & 0.15 & 0.81 \\
& &-20 &0 & - & - & - \\
& & +20&0 & 0.024 & 0.06 & 0.31 \\
& & +10&+10 & - & - & - \\
& &+10 &-10 & - & - & - \\
& & 0&-20 & - & - & - \\
& &0 &+20 & - & - & - \\
\hline
NGC 3079 & CS$(2-1)$ & 0 & 0 & 0.098 & 0.24 & 1.26 \\
& & 0 & -10 & 0.031 & 0.08 & 0.4 \\
& & 0 & +10 & 0.122 & 0.29 & 1.56 \\
& & +10 & 0 & 0.11 & 0.26 & 1.41 \\
& & -10 & 0 & 0.067 & 0.16 & 0.86 \\
& & -10 & +10 & 0.074 & 0.18 & 0.95 \\
& & -10 & +20 & 0.045 & 0.11 & 0.58 \\
\hline\hline

\end{tabular}
\caption*{"-" indicates no fit was possible}
\label{tab:lte}
\end{center}
\end{table*}

\subsection{Non-LTE analysis}

\subsubsection{Chemical modelling}

A time-dependent chemical modelling is necessary in order to take into consideration the history of the gas. Here we are looking at a dense core environment. To determine the properties of the gas in this type of environment, we first complete a detailed chemical modelling, and subsequently link the outputs of this with a molecular line radiative transfer model, as detailed in \citet{2006MNRAS.370..229B}. Since the majority of CS emission is likely to be from dense, star forming regions, we model the formation of CS through the collapse of diffuse molecular gas to a dense core. The chemical model we use is UCL\_CHEM \citep{2004MNRAS.354.1141V}. UCL\_CHEM is a time and depth dependent gas-grain chemical model operating over two phases. Phase I simulates the collapse of initially atomic gas to a dense molecular cloud. During this phase, temperature is a constant at 10 K. Freeze out of species onto grain surfaces occurs; the model uses the chemical network from the UMIST database \citep{2007A&A...466.1197W} comprising of 211 species with 2303 gas and grain surface reactions. Phase II simulates the gas once it has reached a dynamical equilibrium. A burst of star formation is simulated, warming the cloud, increasing the temperature up to a maximum of 400 K. The species included in the model include all extragalactic molecules recorded so far, as well as many galactic species. Surface reactions are mainly based around hydrogenation, allowing saturated species to form. We run a large grid of models varying four parameters. These are, the temperature in phase II (T), the UV radiation field ($\chi$), the cosmic ray ionisation rate ($\zeta$) and the final gas density of the cloud in phase I (n(H$_{2}$)). Each model is ran over a period of \num{e6} years. All other parameters are taken as solar, scaled to the metallicity of the two galaxies where necessary. We ran a large grid of models (summarised in Table ~\ref{tab:uclchemparam}), varying parameters over a wide range, given the lack of prior knowledge of the environment we are considering. By doing this we can see the effect on the chemistry and ultimately the formation of CS in different environments.

\begin{table}[htbp]
\begin{center}
\caption{All permutations of these parameters were ran through UCL\_CHEM}
\label{tab:uclchemparam}
\begin{tabular}{c c c c  }
T (K) & $\chi$ (Habing) & $\zeta$ ($\num{1.3e-17}$ s$^{-1}$) & n(H$_{2}$) (cm$^{-3}$) \\ 
\hline\hline
30 & 1 & 1 & \num{e3} \\
50 & 10 & 10 & \num{e4} \\
100 & 100 & 100 & \num{e5} \\
200 & 1000 & 1000 & \num{e6} \\
300 & 10000 & 10000 & \num{e7}  \\
400 & - & - & \num{e8} \\
\end{tabular}

\end{center}
\end{table}

\subsubsection{Radiative transfer modelling}

The chemical abundances of CS given by UCL\_CHEM are are input to the radiative transfer model SMMOL \citep{2001MNRAS.326.1423R}. This is an accelerated $\Lambda$-iteration (ALI) code for solving multi-level radiative transfer problems. Data on the CS molecule - including energy levels, transition frequencies, Einstein A coefficients and collisional rates - are taken from the LAMDA database \citep{2005A&A...432..369S}. Initially the code calculates the radiation field and the level populations. This is completed under the assumptions of LTE conditions and ISM radiation as input continuum \citep{1994ASPC...58..355B}. The radiation field is then recalculated, a check is made for convergence. This repeats until the model converges. At each radial point the code generates the radiation field and the level populations. The emergent intensity distributions are then convolved with the telescope beam. This means the model directly predicts the line profile for a given source as observed with a given telescope. The code has been successfully benchmarked against other radiative transfer models. The ability to predict a line profile allows us to compare with our observations, giving us a single model that best fits the chemical history and physical conditions of the environment observed. Both SMMOL and UCL\_CHEM are modelling the formation of one dense core. In our beam, there must be many more dense cores. At most, the beam covers an area $\approx$2 kpc (CS$(2-1)$ in NGC 3079) and the typical scale of dense cores are 0.05 pc. It is extremely unlikely CS emission is emanating only from one dense core. If it was, we would not see it as the signal would be too diluted in the beam. It is more likely that the emission is coming from a number of these dense core regions. To account for this, we estimate the number of cores in our beam and scale up the emergent flux by this factor. Our estimate is based on a study by \citet{2005MNRAS.360.1527L}. They estimate that there are 10$^{4}$ hot dense cores in the Milky Way, {and calculate a scale factor for other galaxies based on their star formation rate. We adopt the same methodology, based on the star formation rates of our observed galaxies (see Table ~\ref{tab:galaxies}) and the likely central location of the gas.} Our targets are both starburst galaxies which see a greater rate of star formation than the Milky Way, concentrated over the nuclear bulge of both galaxies \citep{2013ApJ...776...70T, 2010PASJ...62.1085Y, 2006ApJ...649..181S}. As these calculations are only estimates, we scale up by orders of magnitude, and by the same values for both galaxies, although we acknowledge that the regions observed in NGC 3079 are larger than in NGC 6946. {We use scale factors of 1, 10, and 100; leading to our estimates for the number of dense cores to be \num{e4}, \num{e5}, and \num{e6}.} We note that we take the arbitrary end point of the chemical model (\num{e6} years) to input into SMMOL. While we do not claim this to be the age of our cores, we use it as a representative chemical age of the molecular gas at large scales. In fact, time dependent effects may be important in energetic extragalactic environments and have been discussed elsewhere \citep{2013JPCA..117.9593M, 2008ApJ...676..978B}.

Analysis of the models is completed using a $\chi^{2}$ method. We use the equation

\begin{equation}
\chi^{2} = \frac{1}{N} \sum_{i=1}^{N}  \bigg[\frac{F_{\rm mod}(i) - F_{\rm obs}(i)}{F_{\rm obs}(i)}\bigg]^{2},
\end{equation}
where N is the number of lines used, $F_{\rm mod}$ is the integrated flux of the modelled line, and $F_{\rm obs}$ is the integrated flux of the observed line.This formula has been used in previous papers in a similar way (e.g. \citealt{2006MNRAS.370..229B}, \citealt{2004A&A...418.1021D}). We use the difference between the model flux and the observed flux instead of the measurement error because the errors associated with the modelling are greater than those in the observations. This also has the advantage of avoiding giving brighter lines a higher weight. It is possible to use SMMOL to model line profiles of CH$_{3}$OH and H$_{2}$CO. Since neither of these molecules were detected outside of the centre of either galaxy, it was not possible in this study.

\begin{table*}[htbp]

\begin{center}

\caption{Best fit parameters for each location in each galaxy (\num{e6} cores)}
\label{tab:smmolbestfits1}
\begin{tabular}{c r r c c c c c c c}

Galaxy & $\Delta\alpha$ & $\Delta\delta$ & T (K) & $\chi$ (Habing) & $\zeta$ ( $\zeta_{0} = \num{1.7e-17}$ s$^{-1}$) & n (cm$^{-3}$) & $\chi^{2}$ & Notes \\ 
\hline\hline 
NGC 6946 & 0 & 0 & 100 & 1 & 100 & \num{e6} &  0.20  & \\
NGC 6946 & 0 & -10 & 100 & 1 & 1 & \num{e5} &  0.032  & \\
NGC 6946 & 0 & +10 & 100 & 1 & 1 & \num{e5} &  0.046  & \\
NGC 6946 & +10 & 0 & 100 & 1 & 1 & \num{e5} &  0.021  & \\
NGC 6946 & -10 & 0 & 100 & 1 & 1 & \num{e5} &  0.13  & \\
NGC 6946 & -20 & 0 & 100 & 1 & 10 & \num{e7} &  0.55  & poor fit \\
NGC 6946 & +20 & 0 & 300 & 1 & 100 & \num{e7} &  0.06 & $(2-1)$ only \\
NGC 6946 & +10 & +10 & 100 & 1 & 10 & \num{e7} &  1.1 & poor fit \\
NGC 6946 & +10 & -10 & & & & & & no fit  \\
NGC 6946 & 0 & -20 & 400 & 1 & 1000 & \num{e7} &  0.02 & $(2-1)$ only \\
NGC 6946 & 0 & +20 & 100 & 1 & 10 & \num{e7} &  0.80 & poor fit \\
\hline
NGC 3079 & 0 & 0 & 400 & 1 & 10 & \num{e6} & 0.088 & \\
NGC 3079 & +10 & 0 & 400 & 1 & 10 & \num{e6} &  0.009 & $(2-1)$ only \\
NGC 3079 & 0 & +10 & 400 & 1 & 10 & \num{e6} &  0.08 & $(2-1)$ only \\
\end{tabular}

\end{center}
\end{table*}

\begin{table*}[htbp]

\begin{center}

\caption{Best fit parameters for each location in each galaxy (\num{e5} cores)}
\label{tab:smmolbestfits2}
\begin{tabular}{c r r c c c c c c c}

Galaxy & $\Delta\alpha$ & $\Delta\delta$ & T (K) & $\chi$ (Habing) & $\zeta$ ( $\zeta_{0} = \num{1.7e-17}$ s$^{-1}$) & n (cm$^{-3}$) & $\chi^{2}$ & Notes \\ 
\hline\hline 
NGC 6946 & 0 & 0 & 400 & 1 & 10 & \num{e6} &  0.30  & \\
NGC 6946 & 0 & -10 & 200 & 1 & 1 & \num{e6}  & 0.24  & \\
NGC 6946 & 0 & +10 & 100 & 1 & 10 & \num{e5} &  0.25  & \\
NGC 6946 & +10 & 0 & 400 & 1 & 10 & \num{e6} & 0.22  & \\
NGC 6946 & -10 & 0 & 100 & 1 & 10 & \num{e5}  & 0.24  & \\
NGC 6946 & -20 & 0 & 100 & 1 & 10 & \num{e5}  & 0.21  & \\
NGC 6946 & +20 & 0 & 400 & 1 & 10 & \num{e6}  & 0.0001 & $(2-1)$ only \\
NGC 6946 & +10 & +10 & 100 & 1 & 10 & \num{e5}  & 0.31 &  \\
NGC 6946 & +10 & -10 & & & & & & no fit  \\
NGC 6946 & 0 & -20 & 400 & 1 & 10 & \num{e6} &  0.01 & $(2-1)$ only \\
NGC 6946 & 0 & +20 & 400 & 1 & 100 & \num{e5} & 0.02 & $(2-1)$ only \\
\hline
NGC 3079 & 0 & 0 & 400 & 1 & 1 & \num{e5} &  0.37 & \\
NGC 3079 & +10 & 0 & 400 & 1 & 1 & \num{e5} &  0.02 & $(2-1)$ only \\
NGC 3079 & 0 & +10 & 400 & 1 & 1 & \num{e5} &  0.05 & $(2-1)$ only \\
\end{tabular}

\end{center}
\end{table*}

\section{Discussion}

We now analyse the results of our models. It is important to note that while the observations cover a large area, CS emission is predominantly emanating from smaller regions of dense, star forming gas. We assume that there are many of these regions in our beam, based on the work by \citet{2008ApJ...673..183L}. Since we complete this study under the assumption of either \num{e4}, \num{e5} or \num{e6} dense cores in the beam, we discuss each set of models separately. Firstly, under the assumption of \num{e6} cores in the beam, the best fit models for all positions in NGC 3079 are the same (Table ~\ref{tab:smmolbestfits1}). Temperature and density are quite high - 400 K \& \num{e6} cm$^{-3}$, respectively - with a cosmic ray ionisation rate of 100 times the standard rate ($\zeta_{0}$). The high temperature is likely to be somewhat higher than the average temperature across the region we are observing. The high temperature is possibly an indication of the presence of shocks, which we discuss later. Models for NGC 6946 follow the general pattern of decreasing density (\num{e6} cm$^{-3}\rightarrow$ \num{e5} cm$^{-3}$) moving away from the centre of the galaxy (when only including models with good fits to both lines). Models for the outer regions of our observations generally do not provide good fits, or there is not enough data to fit to. Although the lowest $\chi^{2}$s are all for high density (\num{e7} cm$^{-3}$) models. We include these results for completeness, but note that - given the high $\chi^{2}$ values - there are large degeneracies in these models. There are no models with a density \textless \num{e5} cm$^{-3}$ that provide a good fit to the data (i.e. with $\chi^{2}$ \textless 1).

When considering \num{e5} cores in the beam, the best fit models for each position (see Table ~\ref{tab:smmolbestfits2}) in NGC 3079 arise from a different set of parameters from the previous set of models, although again there is no variation across position. Again from a high temperature model, the density and cosmic ray ionisation rate are lower than for the \num{e6} core simulation but do not provide the same quality fit as shown in Table ~\ref{tab:smmolbestfits1}. In contrast with the results for NGC 6946 in Table ~\ref{tab:smmolbestfits1}, the \num{e5} cores models show a higher temperature and higher density for some best fit models, although as with NGC 3079, these do not fit as well as the \num{e6} core models. The UV radiation field has no effect on any of the best fit models for any number of cores, because of the high visual extinction associated with the high density regions where CS emission originates. We note that this does not imply there is a lack of UV radiation but simply that UV is unable to penetrate down to dense cores. We also find that we cannot successfully fit the observations using \num{e4} cores for any region of either galaxy {because the antenna temperature of the theoretical lines is too low to match observations. In many cases, the profiles do not match either.} This possibly provides a lower limit on the number of star forming cores we see within our beam.

\subsection{Shocks}

\begin{figure}[htbp]
\begin{center}
\includegraphics[width=0.85\linewidth, angle=0]{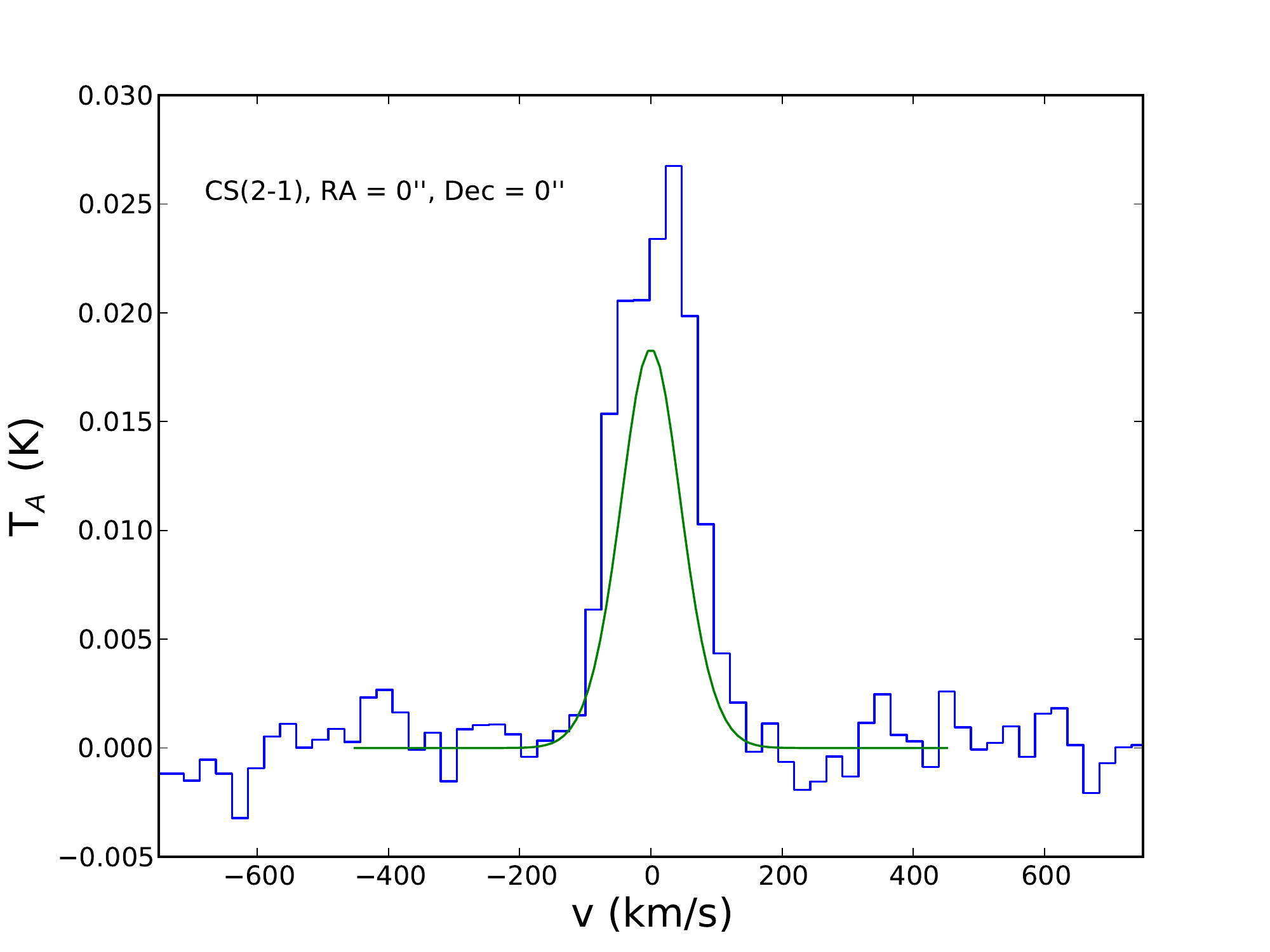}
\includegraphics[width=0.85\linewidth, angle=0]{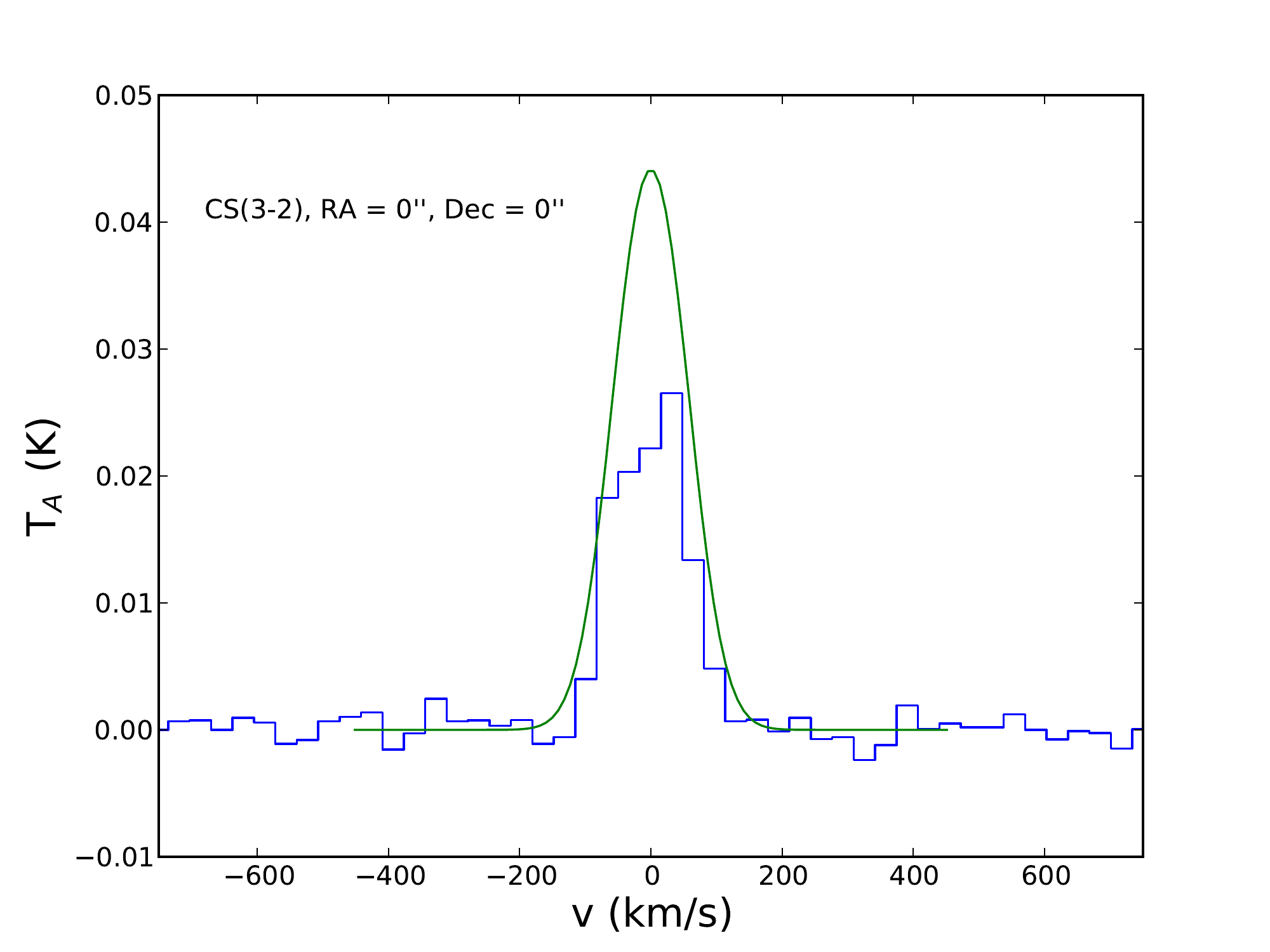}
\caption{Shock model fits for NGC 6946}
\label{fig:shock6946}
\end{center}
\end{figure}

\begin{figure}[htbp]
\begin{center}
\includegraphics[width=0.85\linewidth, angle=0]{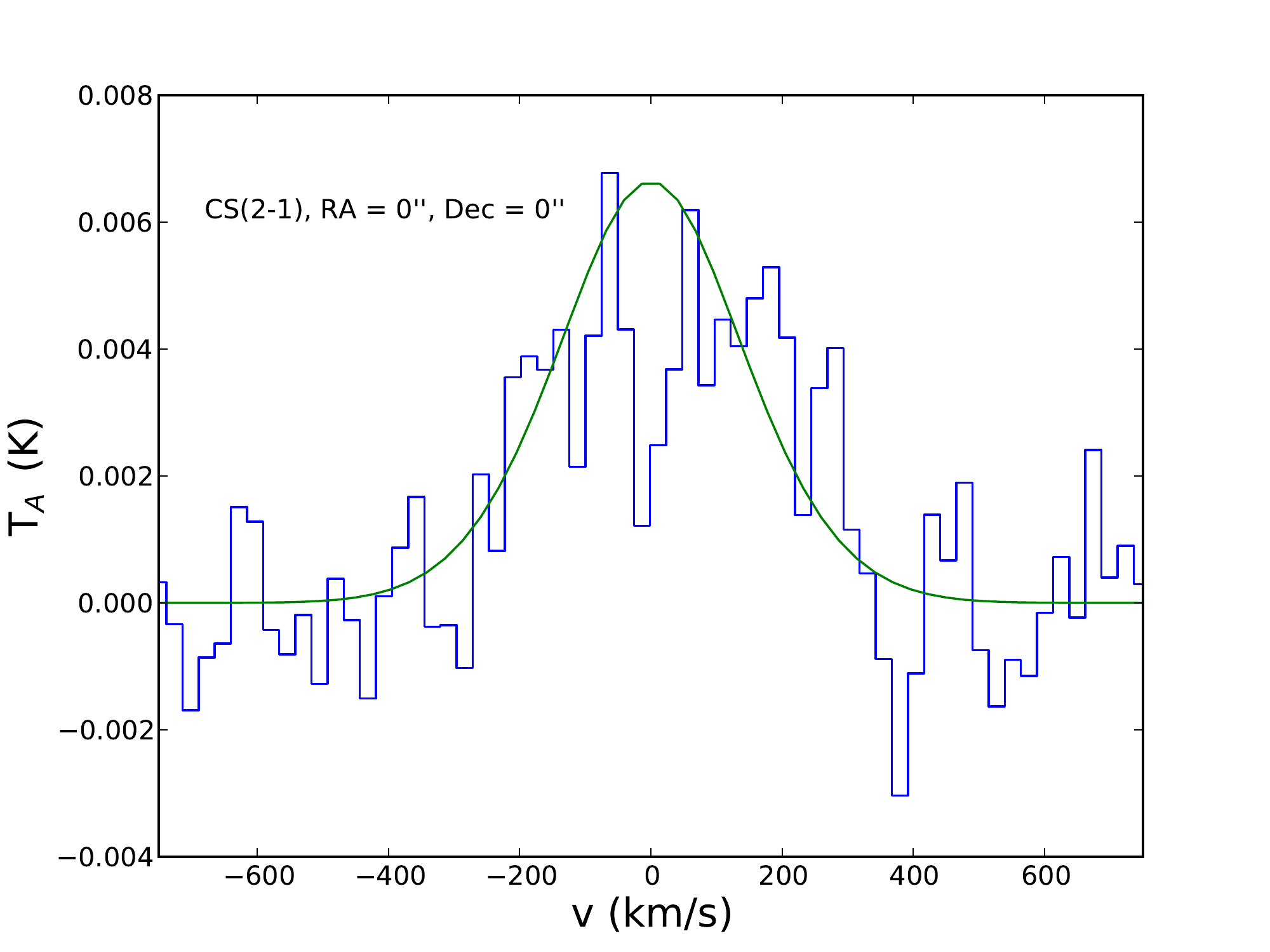}
\includegraphics[width=0.85\linewidth, angle=0]{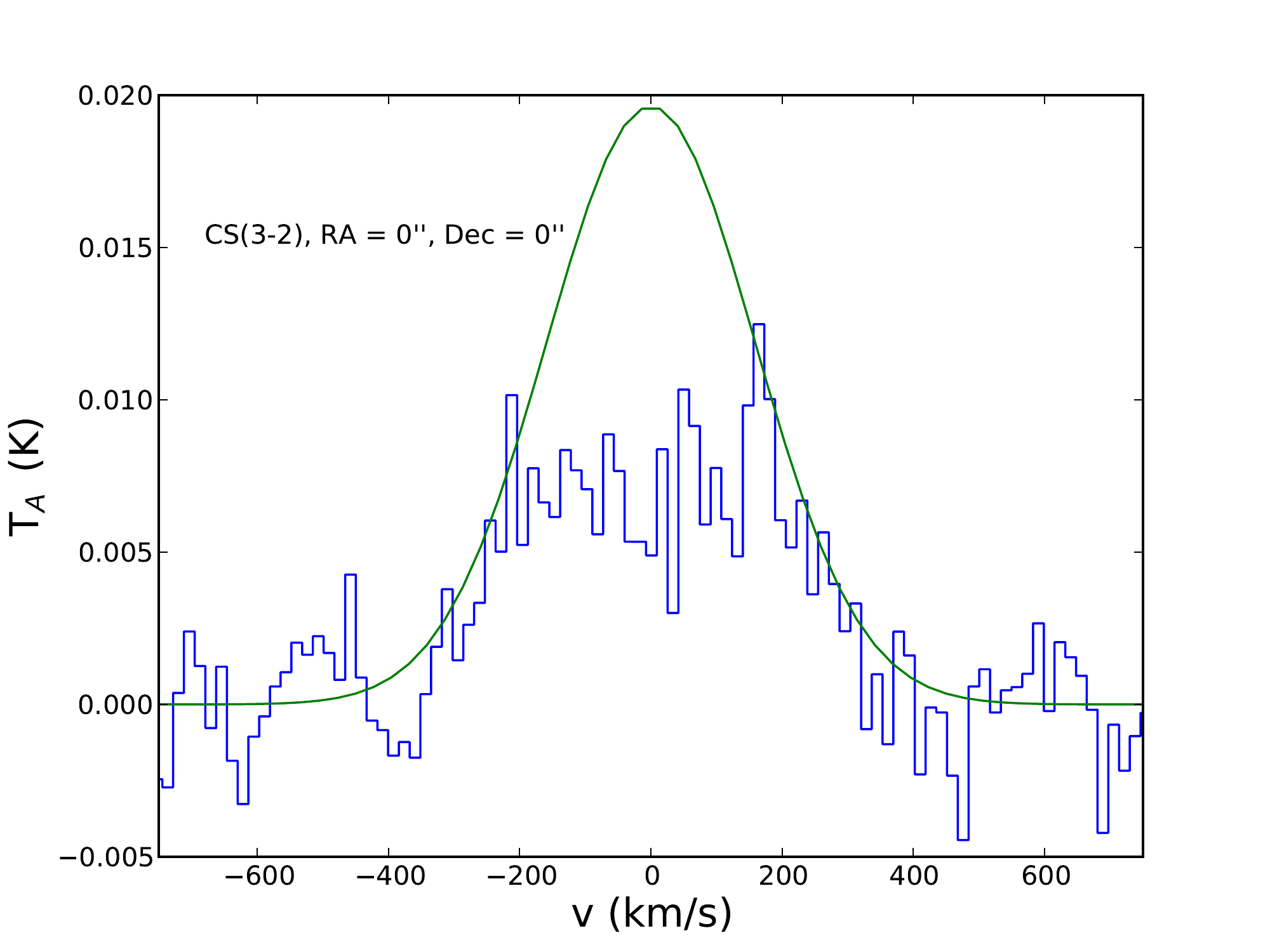}
\caption{Shock model fits for NGC 3079}
\label{fig:shock3079}
\end{center}
\end{figure}

All the discussion above assumes there are no shocks. However, CS has recently been shown to be present in shocked regions (e.g. \citealt{2012MNRAS.419..251N}). This is also indicated by the high temperature models which successfully fit many of the emission regions. Shock temperatures are unlikely to be sustained for the timescale of our models but could be very significant for CS chemistry even over a shorter timescale. UCL\_CHEM is capable of simulating shocks \citep{2001A&A...370.1017V}. We ran shock models for the central position of both galaxies. These were completed with the same best fit parameters as in the \num{e6} core models, as these give the lowest $\chi^{2}$ values. During the shocked period, the temperature rises from 100 K to 500 K, and then reduces back to 100 K. During this period, anything remaining on grain surfaces is sublimated. In our radiative transfer code we then vary the number of cores as before. We see that the best fits to the line intensities of CS$(2-1)$ and CS$(3-2)$ now come from the model assuming that there are \num{e4} dense cores within the beam (see Figures ~\ref{fig:shock6946} \& ~\ref{fig:shock3079}). This suggests that since shocks have a significant effect on CS emission, it is necessary to know whether the region being observed is (or has been) shocked, in order to fully characterise the environment observed. Therefore, observations of shock tracers (such as SiO), would be very useful in conjunction with CS detections.

The line profiles calculated by SMMOL are not just dependent on the physical conditions of the environment, but also on the chemistry of CS. In the following section we analyse the outputs from UCL\_CHEM in order to determine what factors influence CS chemistry.

\subsection{CS formation}
The dominant formation routes to CS are not completely understood. There was initially some agreement \citep{1989ApJ...345..815D, 2002A&A...384.1054L} that most CS formation in the gas phase is a product of the dissociative recombination reaction:

\begin{equation}
\ce{HCS$^{+}$ + e$^{-}$ -> CS + H},
\label{eq:HCS}
\end{equation}

What is less clear is the route to the formation of HCS$^{+}$. \citet{1989ApJ...345..815D} found that the following gas-phase reaction scheme reproduced CS in abundances in the same range as their observations in relatively diffuse regions:

\begin{equation}
\begin{split}
& \ce{S$^{+}$ + CH -> CS$^{+}$ + H}, \\
& \ce{S$^{+}$ + C$_{2}$ -> CS$^{+}$ + C}, \\
& \ce{CS$^{+}$ + H$_{2}$ -> HCS$^{+}$ + H}, \\
\end{split}
\label{eg:routes}
\end{equation}
\\
\\
However, \citet{2002A&A...384.1054L} found that the abundances of S$^{+}$, C$_{2}$, and CH were an order of magnitude lower than observed if the mechanism above was producing the observed quantities of CS. Its high efficiency at low temperatures mean that reaction ~\ref{eq:HCS} is still likely to contribute to CS formation in certain environments. A general, more well agreed upon formation route \citep{2009ApJ...693..804D, 2000A&A...358..257N, 2002A&A...384.1054L} is the substitution reaction:

\begin{equation}
\ce{SO + C -> CS + O},
\label{eq:SO}
\end{equation}

Chemical modelling \citep{1997ApJ...486..316B} has shown that this reaction leads to a high abundance of CS quite swiftly, until formation of CO locks away much of the atomic carbon needed for reaction ~\ref{eq:SO} . At later times, CS abundance drops as  a result of efficient destruction of CS with atomic oxygen:

\begin{equation}
\ce{CS + O -> CO + S},
\label{eq:destruction}
\end{equation}

Routes ~\ref{eq:SO} and ~\ref{eq:destruction} are now believed to dominate the formation and destruction of CS \citep{2002A&A...384.1054L}. Analysis of our chemical models tend to broadly agree with these findings. However, the cosmic ray ionisation rate is an important parameter in determining the dominating mechanisms. A high cosmic ray ionisation rate (such as the one in one of our best fit models for the centre of NGC 6946 - $\zeta =$ 100 $\zeta_{0}$) leads to the destruction of a small percentage of available CO, providing enough atomic carbon to allow reaction ~\ref{eq:SO} to counteract the destruction from reaction ~\ref{eq:destruction}. However, this is also more efficient as there is more atomic oxygen available from CO. At a lower cosmic ray ionisation rate, we find that a significant source of CS destruction is from reaction ~\ref{eq:H3O}. Both the formation and destruction processes are up to two orders of magnitude slower when $\zeta$ is lowered from 100 $\zeta_{0}$ to 1 $\zeta_{0}$:

\begin{equation}
\ce{H$_{3}$O$^{+}$ + CS -> HCS$^{+}$ + H$_{2}$O}, 
\label{eq:H3O}
\end{equation}

During shock treatment the following formation reaction also becomes important:

\begin{equation}
\begin{split}
\ce{C$_{2}$ + S -> CS + C}.
\end{split}
\label{eq:shock}
\end{equation}

Reaction ~\ref{eq:shock} is important during the shock as it both becomes more efficient at high temperatures and the abundance of the reactant species increases. This points to there not being one overwhelmingly dominant way of forming CS. Different reactions are responsible for its formation (and destruction) in different conditions.

\section{Conclusions}

CS emission has been mapped across two nearby starburst galaxies - NGC 3079 and NGC 6946. LTE calculations were undertaken to determine a lower limit to the kinetic temperature in NGC 6946 of 14K. In order to constrain this further, and determine conditions in the dense gas environments being traced by CS, a time dependent gas-grain chemical model (UCL\_CHEM) has been linked to a radiative transfer code (SMMOL) to simulate the line emission. Comparing the observations to the models has led to the conclusion that, given the high temperatures (T = 400 K) often needed for successful fits, models featuring the presence of shocks best reproduce some of our observations. We find that in order to match the observations, emission must be coming from between \num{e4} and \num{e6} dense cores of $\sim$ 0.05 pc. The routes to the formation of CS are  important factors in successfully modelling regions of dense gas. With ever more complex chemical webs, predicted abundances of molecules become highly sensitive to set conditions. We find that shocks provide conditions that are effective in efficiently forming CS in necessary abundances due in part to increased efficiency of Reaction ~\ref{eq:shock}. A high cosmic ray ionisation rate ($\approx$100 $\zeta_{0}$) could also have a significant influence on CS chemistry by increasing the rate of Reaction ~\ref{eq:SO}. In conclusion, CS observations, combined with chemical and radiative transfer modelling are capable of disentangling dense star forming regions from surrounding molecular gas. With better {spatial} resolution it should be possible to quantify further the chemical differentiation across the central regions of starburst galaxies. Future multi-line observations of CS with ALMA would provide the tools to do this.

\section{Acknowledgements}

{We thank the IRAM staff for their help with the observations. This work was supported by the Science \& Technology Facilities Council. We also thank the anonymous referee for their constructive comments, which improved the manuscript.}

\bibliographystyle{aa}
\bibliography{mapping_cs}

\begin{thebibliography}{49}
\expandafter\ifx\csname natexlab\endcsname\relax\def\natexlab#1{#1}\fi

\bibitem[{{Aladro} {et~al.}(2011){Aladro}, {Mart{\'{\i}}n-Pintado},
  {Mart{\'{\i}}n}, {Mauersberger}, \& {Bayet}}]{2011A&A...525A..89A}
{Aladro}, R., {Mart{\'{\i}}n-Pintado}, J., {Mart{\'{\i}}n}, S., {Mauersberger},
  R., \& {Bayet}, E. 2011, A\&A, 525, A89

\bibitem[{{Baan} \& {Irwin}(1995)}]{1995ApJ...446..602B}
{Baan}, W.~A. \& {Irwin}, J.~A. 1995, ApJ, 446, 602

\bibitem[{{Ball} {et~al.}(1985){Ball}, {Sargent}, {Scoville}, {Lo}, \&
  {Scott}}]{1985ApJ...298L..21B}
{Ball}, R., {Sargent}, A.~I., {Scoville}, N.~Z., {Lo}, K.~Y., \& {Scott}, S.~L.
  1985, ApJ L, 298, L21

\bibitem[{{Bayet} {et~al.}(2009){Bayet}, {Aladro}, {Mart{\'{\i}}n}, {Viti}, \&
  {Mart{\'{\i}}n-Pintado}}]{2009ApJ...707..126B}
{Bayet}, E., {Aladro}, R., {Mart{\'{\i}}n}, S., {Viti}, S., \&
  {Mart{\'{\i}}n-Pintado}, J. 2009, ApJ, 707, 126

\bibitem[{{Bayet} {et~al.}(2008{\natexlab{a}}){Bayet}, {Lintott}, {Viti},
  {Mart{\'{\i}}n-Pintado}, {Mart{\'{\i}}n}, {Williams}, \&
  {Rawlings}}]{2008ApJ...685L..35B}
{Bayet}, E., {Lintott}, C., {Viti}, S., {et~al.} 2008{\natexlab{a}}, ApJ L,
  685, L35

\bibitem[{{Bayet} {et~al.}(2008{\natexlab{b}}){Bayet}, {Viti}, {Williams}, \&
  {Rawlings}}]{2008ApJ...676..978B}
{Bayet}, E., {Viti}, S., {Williams}, D.~A., \& {Rawlings}, J.~M.~C.
  2008{\natexlab{b}}, ApJ, 676, 978

\bibitem[{{Bayet} {et~al.}(2011){Bayet}, {Yates}, \&
  {Viti}}]{2011ApJ...728..114B}
{Bayet}, E., {Yates}, J., \& {Viti}, S. 2011, ApJ, 728, 114

\bibitem[{{Benedettini} {et~al.}(2006){Benedettini}, {Yates}, {Viti}, \&
  {Codella}}]{2006MNRAS.370..229B}
{Benedettini}, M., {Yates}, J.~A., {Viti}, S., \& {Codella}, C. 2006, MNRAS,
  370, 229

\bibitem[{{Bergin} \& {Langer}(1997)}]{1997ApJ...486..316B}
{Bergin}, E.~A. \& {Langer}, W.~D. 1997, ApJ, 486, 316

\bibitem[{{Black}(1994)}]{1994ASPC...58..355B}
{Black}, J.~H. 1994, in Astronomical Society of the Pacific Conference Series,
  Vol.~58, The First Symposium on the Infrared Cirrus and Diffuse Interstellar
  Clouds, ed. R.~M. {Cutri} \& W.~B. {Latter}, 355

\bibitem[{{Bronfman} {et~al.}(1996){Bronfman}, {Nyman}, \&
  {May}}]{1996A&AS..115...81B}
{Bronfman}, L., {Nyman}, L.-A., \& {May}, J. 1996, A\&A, 115, 81

\bibitem[{{Crook} {et~al.}(2007){Crook}, {Huchra}, {Martimbeau}, {Masters},
  {Jarrett}, \& {Macri}}]{2007ApJ...655..790C}
{Crook}, A.~C., {Huchra}, J.~P., {Martimbeau}, N., {et~al.} 2007, ApJ, 655, 790

\bibitem[{{Destree} {et~al.}(2009){Destree}, {Snow}, \&
  {Black}}]{2009ApJ...693..804D}
{Destree}, J.~D., {Snow}, T.~P., \& {Black}, J.~H. 2009, ApJ, 693, 804

\bibitem[{{Doty} {et~al.}(2004){Doty}, {Sch{\"o}ier}, \& {van
  Dishoeck}}]{2004A&A...418.1021D}
{Doty}, S.~D., {Sch{\"o}ier}, F.~L., \& {van Dishoeck}, E.~F. 2004, A\&A, 418,
  1021

\bibitem[{{Drdla} {et~al.}(1989){Drdla}, {Knapp}, \& {van
  Dishoeck}}]{1989ApJ...345..815D}
{Drdla}, K., {Knapp}, G.~R., \& {van Dishoeck}, E.~F. 1989, ApJ, 345, 815

\bibitem[{{Engelbracht} {et~al.}(1996){Engelbracht}, {Rieke}, {Rieke}, \&
  {Latter}}]{1996ApJ...467..227E}
{Engelbracht}, C.~W., {Rieke}, M.~J., {Rieke}, G.~H., \& {Latter}, W.~B. 1996,
  ApJ, 467, 227

\bibitem[{{Gao} \& {Solomon}(2004)}]{2004ApJ...606..271G}
{Gao}, Y. \& {Solomon}, P.~M. 2004, \apj, 606, 271

\bibitem[{{Goldsmith} \& {Langer}(1999)}]{1999ApJ...517..209G}
{Goldsmith}, P.~F. \& {Langer}, W.~D. 1999, ApJ, 517, 209

\bibitem[{{Irwin} \& {Seaquist}(1991)}]{1991ApJ...371..111I}
{Irwin}, J.~A. \& {Seaquist}, E.~R. 1991, ApJ, 371, 111

\bibitem[{{Kepley} {et~al.}(2014){Kepley}, {Leroy}, {Frayer}, {Usero},
  {Marvil}, \& {Walter}}]{2014ApJ...780L..13K}
{Kepley}, A.~A., {Leroy}, A.~K., {Frayer}, D., {et~al.} 2014, \apjl, 780, L13

\bibitem[{{Koda} {et~al.}(2002){Koda}, {Sofue}, {Kohno}, {Nakanishi},
  {Onodera}, {Okumura}, \& {Irwin}}]{2002ApJ...573..105K}
{Koda}, J., {Sofue}, Y., {Kohno}, K., {et~al.} 2002, ApJ, 573, 105

\bibitem[{{Leroy} {et~al.}(2008){Leroy}, {Walter}, {Brinks}, {Bigiel}, {de
  Blok}, {Madore}, \& {Thornley}}]{2008AJ....136.2782L}
{Leroy}, A.~K., {Walter}, F., {Brinks}, E., {et~al.} 2008, ApJ, 136, 2782

\bibitem[{{Leroy} {et~al.}(2011){Leroy}, {Walter}, {Schruba}, {Bigiel},
  {Foyle}, \& {HERACLES Team}}]{2011AAS...21724614L}
{Leroy}, A.~K., {Walter}, F., {Schruba}, A., {et~al.} 2011, in Bulletin of the
  American Astronomical Society, Vol.~43, American Astronomical Society Meeting
  Abstracts 217, 246.14

\bibitem[{{Levine} {et~al.}(2008){Levine}, {Helfer}, {Meijerink}, \&
  {Blitz}}]{2008ApJ...673..183L}
{Levine}, E.~S., {Helfer}, T.~T., {Meijerink}, R., \& {Blitz}, L. 2008, ApJ,
  673, 183

\bibitem[{{Lintott} {et~al.}(2005){Lintott}, {Viti}, {Williams}, {Rawlings}, \&
  {Ferreras}}]{2005MNRAS.360.1527L}
{Lintott}, C.~J., {Viti}, S., {Williams}, D.~A., {Rawlings}, J.~M.~C., \&
  {Ferreras}, I. 2005, MNRAS, 360, 1527

\bibitem[{{Lucas} \& {Liszt}(2002)}]{2002A&A...384.1054L}
{Lucas}, R. \& {Liszt}, H.~S. 2002, A\&A, 384, 1054

\bibitem[{{Mart{\'{\i}}n} {et~al.}(2005){Mart{\'{\i}}n},
  {Mart{\'{\i}}n-Pintado}, {Mauersberger}, {Henkel}, \&
  {Garc{\'{\i}}a-Burillo}}]{2005ApJ...620..210M}
{Mart{\'{\i}}n}, S., {Mart{\'{\i}}n-Pintado}, J., {Mauersberger}, R., {Henkel},
  C., \& {Garc{\'{\i}}a-Burillo}, S. 2005, ApJ, 620, 210

\bibitem[{{Mart{\'{\i}}n} {et~al.}(2006){Mart{\'{\i}}n}, {Mauersberger},
  {Mart{\'{\i}}n-Pintado}, {Henkel}, \&
  {Garc{\'{\i}}a-Burillo}}]{2006ApJS..164..450M}
{Mart{\'{\i}}n}, S., {Mauersberger}, R., {Mart{\'{\i}}n-Pintado}, J., {Henkel},
  C., \& {Garc{\'{\i}}a-Burillo}, S. 2006, ApJ, 164, 450

\bibitem[{{Mauersberger} {et~al.}(1989){Mauersberger}, {Henkel}, {Wilson}, \&
  {Harju}}]{1989A&A...226L...5M}
{Mauersberger}, R., {Henkel}, C., {Wilson}, T.~L., \& {Harju}, J. 1989, A\&A,
  226, L5

\bibitem[{{Meijerink} {et~al.}(2013){Meijerink}, {Spaans}, {Kamp}, {Aresu},
  {Thi}, \& {Woitke}}]{2013JPCA..117.9593M}
{Meijerink}, R., {Spaans}, M., {Kamp}, I., {et~al.} 2013, Journal of Physical
  Chemistry A, 117, 9593

\bibitem[{{Mineo} {et~al.}(2014){Mineo}, {Gilfanov}, {Lehmer}, {Morrison}, \&
  {Sunyaev}}]{2014MNRAS.437.1698M}
{Mineo}, S., {Gilfanov}, M., {Lehmer}, B.~D., {Morrison}, G.~E., \& {Sunyaev},
  R. 2014, MNRAS, 437, 1698

\bibitem[{{Nicholas} {et~al.}(2012){Nicholas}, {Rowell}, {Burton}, {Walsh},
  {Fukui}, {Kawamura}, \& {Maxted}}]{2012MNRAS.419..251N}
{Nicholas}, B.~P., {Rowell}, G., {Burton}, M.~G., {et~al.} 2012, MNRAS, 419,
  251

\bibitem[{{Nilsson} {et~al.}(2000){Nilsson}, {Hjalmarson}, {Bergman}, \&
  {Millar}}]{2000A&A...358..257N}
{Nilsson}, A., {Hjalmarson}, {\AA}., {Bergman}, P., \& {Millar}, T.~J. 2000,
  A\&A, 358, 257

\bibitem[{{Onishi} {et~al.}(1998){Onishi}, {Mizuno}, {Kawamura}, {Ogawa}, \&
  {Fukui}}]{1998ApJ...502..296O}
{Onishi}, T., {Mizuno}, A., {Kawamura}, A., {Ogawa}, H., \& {Fukui}, Y. 1998,
  \apj, 502, 296

\bibitem[{{Rawlings} \& {Yates}(2001)}]{2001MNRAS.326.1423R}
{Rawlings}, J.~M.~C. \& {Yates}, J.~A. 2001, MNRAS, 326, 1423

\bibitem[{{Schinnerer} {et~al.}(2007){Schinnerer}, {B{\"o}ker}, {Emsellem}, \&
  {Downes}}]{2007A&A...462L..27S}
{Schinnerer}, E., {B{\"o}ker}, T., {Emsellem}, E., \& {Downes}, D. 2007, \aap,
  462, L27

\bibitem[{{Schinnerer} {et~al.}(2006){Schinnerer}, {B{\"o}ker}, {Emsellem}, \&
  {Lisenfeld}}]{2006ApJ...649..181S}
{Schinnerer}, E., {B{\"o}ker}, T., {Emsellem}, E., \& {Lisenfeld}, U. 2006,
  ApJ, 649, 181

\bibitem[{{Sch{\"o}ier} {et~al.}(2005){Sch{\"o}ier}, {van der Tak}, {van
  Dishoeck}, \& {Black}}]{2005A&A...432..369S}
{Sch{\"o}ier}, F.~L., {van der Tak}, F.~F.~S., {van Dishoeck}, E.~F., \&
  {Black}, J.~H. 2005, A\&A, 432, 369

\bibitem[{{Skrutskie} {et~al.}(2006){Skrutskie}, {Cutri}, {Stiening},
  {Weinberg}, {Schneider}, {Carpenter}, {Beichman}, {Capps}, {Chester},
  {Elias}, {Huchra}, {Liebert}, {Lonsdale}, {Monet}, {Price}, {Seitzer},
  {Jarrett}, {Kirkpatrick}, {Gizis}, {Howard}, {Evans}, {Fowler}, {Fullmer},
  {Hurt}, {Light}, {Kopan}, {Marsh}, {McCallon}, {Tam}, {Van Dyk}, \&
  {Wheelock}}]{2006AJ....131.1163S}
{Skrutskie}, M.~F., {Cutri}, R.~M., {Stiening}, R., {et~al.} 2006, ApJ, 131,
  1163

\bibitem[{{Sliwa} {et~al.}(2014){Sliwa}, {Wilson}, {Iono}, {Peck}, \&
  {Matsushita}}]{2014ApJ...796L..15S}
{Sliwa}, K., {Wilson}, C.~D., {Iono}, D., {Peck}, A., \& {Matsushita}, S. 2014,
  \apjl, 796, L15

\bibitem[{{Tsai} {et~al.}(2013){Tsai}, {Turner}, {Beck}, {Meier}, \&
  {Wright}}]{2013ApJ...776...70T}
{Tsai}, C.-W., {Turner}, J.~L., {Beck}, S.~C., {Meier}, D.~S., \& {Wright},
  S.~A. 2013, ApJ, 776, 70

\bibitem[{{Turner} \& {Ho}(1983)}]{1983ApJ...268L..79T}
{Turner}, J.~L. \& {Ho}, P.~T.~P. 1983, ApjL, 268, L79

\bibitem[{{Veilleux} {et~al.}(1994){Veilleux}, {Cecil}, {Bland-Hawthorn},
  {Tully}, {Filippenko}, \& {Sargent}}]{1994ApJ...433...48V}
{Veilleux}, S., {Cecil}, G., {Bland-Hawthorn}, J., {et~al.} 1994, ApJ, 433, 48

\bibitem[{{Viti} {et~al.}(2001){Viti}, {Caselli}, {Hartquist}, \&
  {Williams}}]{2001A&A...370.1017V}
{Viti}, S., {Caselli}, P., {Hartquist}, T.~W., \& {Williams}, D.~A. 2001, A\&A,
  370, 1017

\bibitem[{{Viti} {et~al.}(2004){Viti}, {Collings}, {Dever}, {McCoustra}, \&
  {Williams}}]{2004MNRAS.354.1141V}
{Viti}, S., {Collings}, M.~P., {Dever}, J.~W., {McCoustra}, M.~R.~S., \&
  {Williams}, D.~A. 2004, MNRAS, 354, 1141

\bibitem[{{Woodall} {et~al.}(2007){Woodall}, {Ag{\'u}ndez}, {Markwick-Kemper},
  \& {Millar}}]{2007A&A...466.1197W}
{Woodall}, J., {Ag{\'u}ndez}, M., {Markwick-Kemper}, A.~J., \& {Millar}, T.~J.
  2007, A\&A, 466, 1197

\bibitem[{{Yamagishi} {et~al.}(2010){Yamagishi}, {Kaneda}, {Ishihara},
  {Komugi}, {Suzuki}, \& {Onaka}}]{2010PASJ...62.1085Y}
{Yamagishi}, M., {Kaneda}, H., {Ishihara}, D., {et~al.} 2010, PASJ, 62, 1085

\bibitem[{{Zhang} {et~al.}(2014){Zhang}, {Gao}, {Henkel}, {Zhao}, {Wang},
  {Menten}, \& {G{\"u}sten}}]{2014ApJ...784L..31Z}
{Zhang}, Z.-Y., {Gao}, Y., {Henkel}, C., {et~al.} 2014, ApJL, 784, L31

\bibitem[{{Zinnecker} \& {Yorke}(2007)}]{2007ARA&A..45..481Z}
{Zinnecker}, H. \& {Yorke}, H.~W. 2007, A\&AR, 45, 481

\end{thebibliography}

\end{document}